\documentclass[twocolumn,showpacs,preprintnumbers,amsmath,amssymb,superscriptaddress]{revtex4}

\usepackage{graphicx}
\usepackage{dcolumn}
\usepackage{bm}
\usepackage{amsmath}
\usepackage{amssymb}
\usepackage{latexsym}
\usepackage{setspace}
\usepackage{graphics}
\usepackage{amsfonts}
\usepackage{epsfig}
\def\lsim{\mathrel{\rlap{\lower4pt\hbox{\hskip1pt$\sim$}}
    \raise1pt\hbox{$<$}}}         
\def\gsim{\mathrel{\rlap{\lower4pt\hbox{\hskip1pt$\sim$}}
    \raise1pt\hbox{$>$}}}         

\begin{document}


\title{Fluctuation modes of a twist-bend nematic liquid crystal\\}

\author{Z. Parsouzi.A.Sh}
\affiliation{%
Department of Physics, Kent State University, Kent, Ohio 44242, USA\\
}%

\author{S. M. Shamid}
\affiliation{%
Department of Physics, Kent State University, Kent, Ohio 44242, USA\\
}%

\author{V. Borshch}
\affiliation{%
Chemical Physics Interdisciplinary Program and Liquid Crystal Institute, Kent State University, Kent, Ohio 44242, USA\\
}%

\author{P. K. Challa}
\affiliation{%
Department of Physics, Kent State University, Kent, Ohio 44242, USA\\
}%

\author{M. G. Tamba}
\affiliation{%
Department of Chemistry, University of Hull, Hull HU6 7RX, UK\\
}%

\author{C. Welch}
\affiliation{%
Department of Chemistry, University of Hull, Hull HU6 7RX, UK\\
}%

\author{G. H. Mehl}
\affiliation{%
Department of Chemistry, University of Hull, Hull HU6 7RX, UK\\
}%

\author{J. T. Gleeson}
\affiliation{%
Department of Physics, Kent State University, Kent, Ohio 44242, USA\\
}%

\author{A. Jakli}
\affiliation{%
Chemical Physics Interdisciplinary Program and Liquid Crystal Institute, Kent State University, Kent, Ohio 44242, USA\\
}%

\author{O. D. Lavrentovich}
\affiliation{%
Chemical Physics Interdisciplinary Program and Liquid Crystal Institute, Kent State University, Kent, Ohio 44242, USA\\
}%

\author{D. W. Allender}
\affiliation{%
Department of Physics, Kent State University, Kent, Ohio 44242, USA\\
}%

\author{J. V. Selinger}
\affiliation{%
Chemical Physics Interdisciplinary Program and Liquid Crystal Institute, Kent State University, Kent, Ohio 44242, USA\\
}%

\author{S. Sprunt}
\email{ssprunt@kent.edu}
\affiliation{%
Department of Physics, Kent State University, Kent, Ohio 44242, USA\\
}%

\date{\today}

\begin{abstract}
We report a dynamic light scattering study of the fluctuation modes in a thermotropic liquid crystalline mixture of monomer and dimer compounds that exhibits the twist-bend nematic ($\mathrm{N_{TB}}$) phase. The results reveal a spectrum of overdamped fluctuations that includes two nonhydrodynamic and one hydrodynamic mode in the $\mathrm{N_{TB}}$ phase, and a single nonhydrodynamic plus two hydrodynamic modes (the usual nematic optic axis or director fluctuations) in the higher temperature, uniaxial nematic phase. The properties of these fluctuations and the conditions for their observation are comprehensively explained by a Landau-deGennes expansion of the free energy density in terms of heliconical director and helical polarization fields that characterize the $\mathrm{N_{TB}}$ structure, with the latter serving as the primary order parameter. A ``coarse-graining" approximation simplifies the theoretical analysis, and enables us to demonstrate quantitative agreement between the calculated and experimentally determined temperature dependence of the mode relaxation rates. 
\end{abstract}


\maketitle

\section{Introduction}
The twist-bend nematic ($\mathrm{N_{TB}}$) phase is a fascinating new addition to the family of orientationally-ordered, liquid crystalline states of matter. It has been described as the ``fifth nematic phase" \cite{Chen_PNAS}, complementing the uniaxial, biaxial, chiral helical (cholesteric), and blue phase nematics. Originally proposed by Meyer \cite{Meyer}, and later elaborated on theoretically by Dozov \cite{Dozov}, the existence of the $\mathrm{N_{TB}}$ phase was suggested experimentally \cite{Cestari_PRE} and subsequently confirmed \cite{Chen_PNAS,Borshch_Nature} in low molecular weight liquid crystals (LCs) containing achiral dimers having an odd-numbered hydrocarbon linkage between the mesogenic ends. Interest in these materials was also inspired by simulation studies \cite{Luckhurst}, which predicted a nematic--nematic transition in LC dimers with odd-numbered linkages. 

The $\mathrm{N_{TB}}$ state possesses some remarkable properties. First, the average molecular long axis (specified by a unit vector $\hat{\mathbf{n}}$ called the director) simultaneously bends and twists in space. In the case of LC dimers with odd linkage, the specific tendency to bend is presumably caused by an all-trans conformation of the molecules, which results in their having a pronounced bent shape. The addition of twist allows the bend to be uniform everywhere in space. The combination of bend and twist produces an oblique helicoidal (or heliconical) winding of the director (Fig. 1), with a cone angle $\beta$ (angle between the $\hat{\mathbf{n}}$ and the helicoidal axis) of magnitude $\simeq 10^\circ$. This differs from an ordinary cholesteric LC phase, where a pure twist of $\hat{\mathbf{n}}$ results in a right-angle helicoid ($\beta = 90^\circ$). 

Second, the helicoidal pitch in the $\mathrm{N_{TB}}$ phase is on a molecular scale -- i.e., on the order of 10~nm \cite{Chen_PNAS,Borshch_Nature} -- compared with cholesterics, where the supramolecular pitch typically exceeds 100~nm. The much larger pitch of a cholesteric may be attributed to the relative freedom of rotations around the long molecular axes, when the latter are orthogonal to the helical axis ($\beta=90^\circ$). This configuration mitigates the chiral part of intermolecular interactions \cite{Kamien}. By contrast, in the $\mathrm{N_{TB}}$ state (with $\beta < 90^\circ$), the bend-imposed hindrance of molecular rotations results in a much shorter, nanoscale modulation, which, however, remains purely orientational in nature -- i.e, there is no associated variation in mass density (no Bragg peak detected by X-ray scattering \cite{Chen_PNAS,Borshch_Nature,Cestari_PRE}).

Third, and again unlike a cholesteric, the component molecules of $\mathrm{N_{TB}}$-forming LCs are typically achiral. Thus, the chiral nature of the helicoidal structure is spontaneously generated, with degenerate domains of left- and right-handed helicity.

Finally, although the $\mathrm{N_{TB}}$ phase shows no evidence of a macroscopic polarization, the flexoelectric effect \cite{BMeyer} associated with spontaneous bending of $\hat{\mathbf{n}}$ and the recent observation of an electroclinic effect \cite{CMeyer_PRL} in the $\mathrm{N_{TB}}$ phase suggest that a short-pitch helical polarization field is tied to the heliconical director structure (see Fig.~1). A recent theory \cite{Shamid_PRE} describing the transition between uniaxial and twist-bend nematic phases invokes such a polarization field as the primary order parameter. 

Despite the intense experimental and theoretical efforts to explore the $\mathrm{N_{TB}}$ phase, the nature of collective fluctuation modes associated with the short-pitch helicoidal structure remains an open question. It is a vital one to address, since the spectrum and dispersion of these modes are closely related to the basic structural features and to the relevant order parameter(s), and because properties of the fluctuations provide an important test of theories describing the formation of the $\mathrm{N_{TB}}$ state. Although previous dynamic light scattering measurements \cite{Adlem_PRE} revealed a softening of the elastic constant associated with bend distortions of the director above the N-$\mathrm{N_{TB}}$ transition, they did not probe fluctuation modes specifically associated with the heliconical $N_{TB}$ structure. Here we report, to the best of our our knowledge, the first DLS study of fluctuations within the $\mathrm{N_{TB}}$ phase and their critical behavior near the transition. Our measurements reveal a pair of strongly temperature-dependent nonhydrodynamic modes plus a single hydrodynamic mode in the $\mathrm{N_{TB}}$ phase, and a single nonhydrodynamic mode and pair of hydrodynamic modes (the usual director modes of a uniaxial nematic) in the higher temperature nematic phase. We demonstrate excellent agreement between the behavior of the observed modes and new theoretical predictions based on a ``coarse-grained" version of a Landau-de Gennes free energy for the nematic to $\mathrm{N_{TB}}$ transition \cite{Shamid_PRE}. 

The coarse-graining approximation, inspired in part by earlier theoretical work on cholesterics \cite{Lubensky} and appropriate in the limit of helical pitch much shorter than an optical wavelength, treats surfaces of constant phase in the heliconical structure as ``pseudo-layers." Within this approximation, which has been previously used to explain the effect of high magnetic fields on the $\mathrm{N_{TB}}$ phase \cite{Challa_PRE} and to account for its flow properties \cite{Salili_RSC}, the normal fluctuation modes involving the director may be mapped onto those of a chiral smectic-A phase, with effective layer spacing equal to the pitch, effective director parallel to the local pitch axis, and effective elastic constants that arise from the short-pitch orientational modulation rather than from a true mass density wave. 

An alternative approach to coarse-graining the $\mathrm{N_{TB}}$ phase has recently been published \cite{CMeyer_SoftMatter}. Our theory is generally consistent with that work, in that both theories describe the coarse-grained $\mathrm{N_{TB}}$ phase as an effective chiral smectic-A phase, with elastic constants for layer compression and layer bending. The new aspects of our approach are that it describes nonhydrodynamic as well as hydrodynamic fluctuation modes, and it relates all of the modes to microscopic fluctuations of the polarization as well as the director field. Our experimental results agree well with this theoretical approach and, perhaps more significantly, support the centrality of a helical polarization field in describing the nematic to $\mathrm{N_{TB}}$ transition -- an aspect which fundamentally distinguishes the $\mathrm{N_{TB}}$ phase from the other known nematic LC states, including, in particular, the cholesteric phase.

The body of this paper is organized as follows: In Sec.~II, we provide essential details about the experimental setup and procedures, while Sec.~III describes the key experimental results. Sec.~IV presents a detailed discussion of a Landau theory for a N-$\mathrm{N_{TB}}$ transition and the coarse-graining approach to calculate the normal fluctuation modes associated with the twist-bend structure. The theoretical predictions are compared to the experimental results in Sec.~V, and Sec.~VI summarizes our findings and offers some concluding remarks.

\section{Experimental details}    
DLS measurements were performed on a 30/70 wt\% mixture of the monomer and dimer compounds shown in Fig.~2 \cite{Mehl}. This mixture has the phase sequence isotropic~$\rightarrow$~(uniaxial)~nematic~(N)~$\rightarrow$~$\mathrm{N_{TB}}$~$\rightarrow$~crystal in cooling, with N to $\mathrm{N_{TB}}$ transition temperature, $T_{TB} = 94.2^\circ$C (measured with a calibrated platinum RTD in our light scattering oven). The $\mathrm{N_{TB}}$ phase in this system has been characterized by a variety of techniques \cite{Borshch_Nature}; for our purposes, its choice afforded the possibility to obtain high quality alignment of the average director (optic axis) in either homogeneous planar or homeotropic configurations -- i.e., with average $\hat{\mathbf{n}}$ parallel or normal to the plane of the optical substrates, respectively -- using thin ($5~\mu$m) cells with appropriate surface treatments. 

\begin{figure}[tbp]
\centering
\includegraphics[width=0.9\columnwidth]{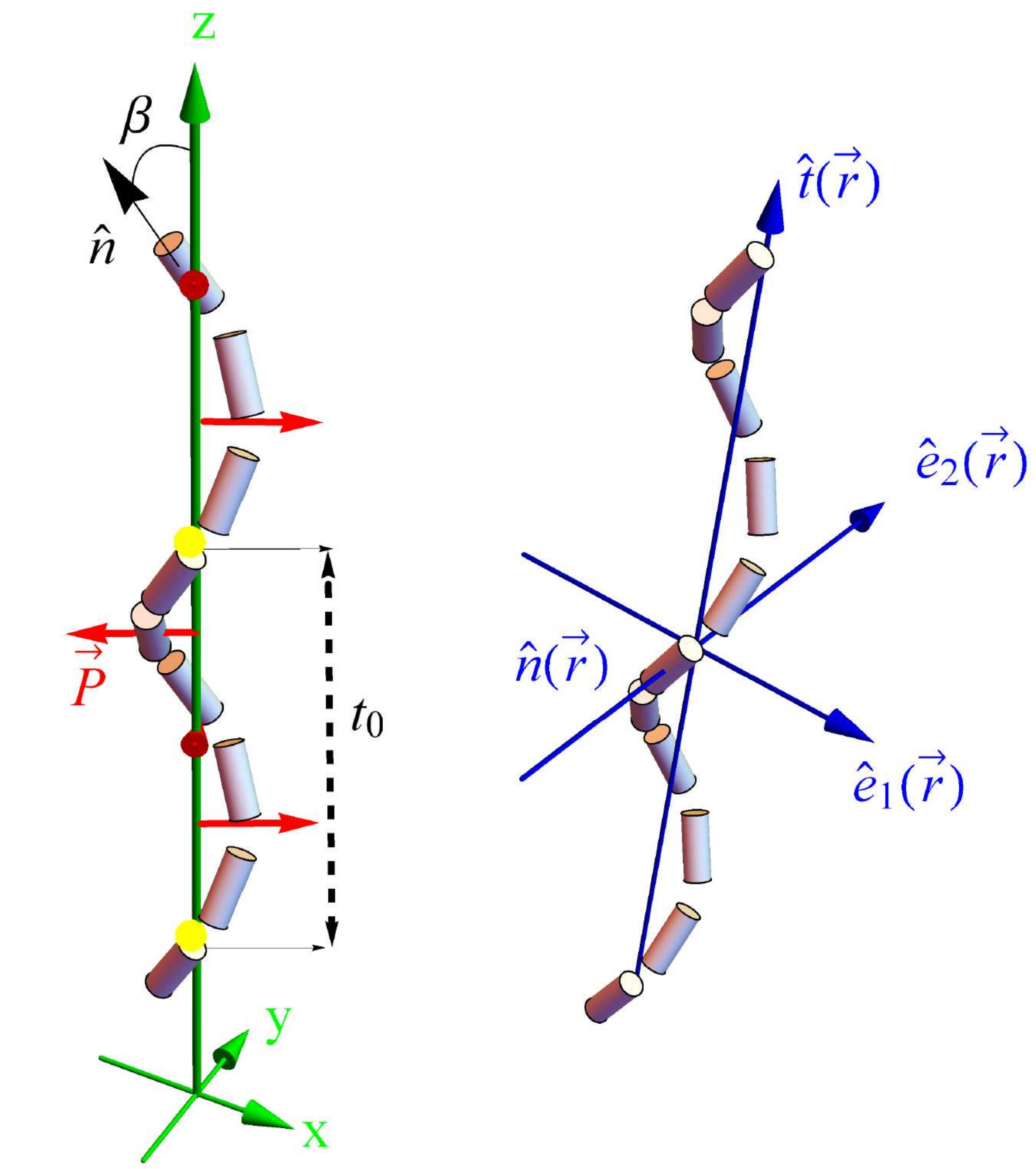}
\caption{(Color online) Left: Schematic representation of the $\mathrm{N_{TB}}$ phase structure, showing heliconical director $\hat{\mathbf{n}}$ (with cone angle $\beta$ and helical pitch $t_0$) and helical polarization field $\mathbf{P}$. Right: Frame of reference used to describe spatial variations of the average director or pitch axis, $\hat{\mathbf{t}}$, on length scales much longer than the pitch (see Theory section). The orthogonal unit vectors $\hat{\mathbf{e}}_1$ and $\hat{\mathbf{e}}_2$ form a right-handed system with $\hat{\mathbf{t}}$. The $xyz$ axes are fixed in the laboratory frame.}
\end{figure}

Our DLS measurements utilized two depolarized scattering geometries -- G1 and G2, depicted in Fig.~2 -- in which homodyne time correlation functions of the depolarized scattered intensity of laser light (wavelength $\lambda=532$~nm) are collected as a function of scattering vector $\mathbf{q}$ and temperature $T$. 

In geometry G1 (Fig.~2), the average director is planar-aligned and oriented perpendicular to the scattering plane. We set the wavevector $\mathbf{k}_i$ of the incident beam to an angle $\theta_i = 0^\circ$ (measured with respect to the substrate normal) and varied the direction of wavevector $\mathbf{k}_s$ of the scattered light (described by scattering angle $\theta_s$ relative to the substrate normal). In the nematic phase, for large $\theta_s$, this geometry probes nearly pure splay fluctuations of the director with relaxation rate $\Gamma_1^n \sim q^2$.

In geometry G2, the average director is parallel to the substrate normal (homeotropic alignment) and lies in the scattering plane; in this case, depolarized DLS in the nematic phase probes a combination of overdamped twist and bend fluctuations of $\hat{\mathbf{n}}$ -- the hydrodynamic twist-bend director mode, with relaxation rate $\Gamma_2^n \sim q^2$. The incident wavevector $\mathbf{k}_i$ was fixed at $\theta_i = 15^\circ$ or $35^\circ$, while the direction of $\mathbf{k}_s$ was varied between $\theta_s = -10^\circ$ and $50^\circ$, with respect to average $\hat{\mathbf{n}}$. When $\theta_s = 0^\circ$, $\mathbf{k}_s$ lies along $\langle \hat{\mathbf{n}} \rangle$, and the scattering from director fluctuations is nominally extinguished (``dark director" geometry). This choice of $\theta_s$ provides an opportunity to detect fluctuation modes that do not originate from $\hat{\mathbf{n}}$ and contribute to the dielectric tensor in their own right.

\begin{figure}[tbp]
\includegraphics[width=1.0\columnwidth]{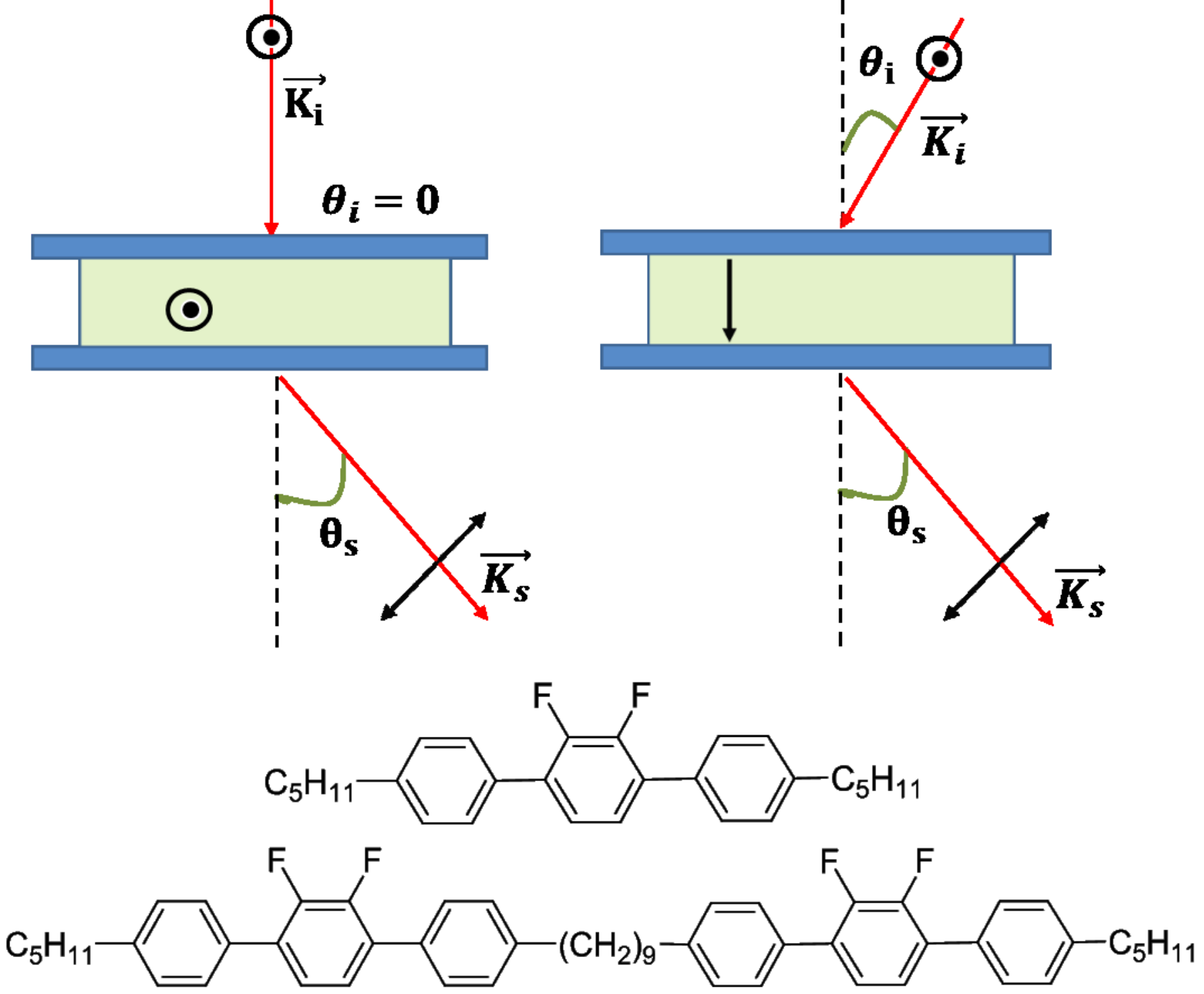}
\caption{(Color online) Top: Light scattering geometries G1 (left) and G2 (right) described in the text, with the average director (optic axis) in the sample cell indicated by the arrow pointing out of the page for G1 (homogeneous planar alignment with average $\hat{\mathbf{n}}$ normal to the scattering plane) or the downward arrow for G2 (homeotropic alignment with average $\hat{\mathbf{n}}$ in the plane). The orientations of polarizer and analyzer are similarly indicated. Bottom: Chemical structure of the monomer and dimer compounds utilized for the present study. The 30/70 wt\% mixture exhibits a N--$\mathrm{N_{TB}}$ phase transition at $94.2^\circ$C.}
\end{figure}

\section{Results}
Fig.~3 shows polarizing microscope images of a homeotropic sample of the mixture during the uniaxial nematic to twist-bend (N--$\mathrm{N_{TB}}$) transition, with the lower left part of each picture corresponding to the N and the upper right part to the $\mathrm{N_{TB}}$ phase. Fig.~3(a) confirms the high quality of the homeotropic alignment of the average director and its persistence across the N--$\mathrm{N_{TB}}$ transition. In the $\mathrm{N_{TB}}$ phase, the average value of $\hat{\mathbf{n}}$ is the pitch axis $\hat{\mathbf{t}}$ of the heliconical structure, which is oriented perpendicular to the substrates (image plane in the figure). Under an applied AC voltage (5 V @ 10 KHz), a second order Freedericsz transition (reorientation of the average director in the center of the sample) is observed in the N region, while the $\mathrm{N_{TB}}$ region is unchanged, Figure 3(b). In the $\mathrm{N_{TB}}$ region, the reorientation occurs at higher voltage (7 V @ 10 KHz, Fig.~3(c)), and in the form of propagating focal-conic domains (FCDs), such as is usually observed in smectic liquid crystals \cite{Lavrentovich_PRL,Jakli_APL}. The ``pseudo-layered" nature of the heliconical structure \cite{Challa_PRE} is reflected in the gradual relaxation of the FCDs to homeotropic alignment after removal of the field. As Fig. 3(d) indicates, the slow relaxation rate and presence of FCDs are quite distinct from the behavior observed in the nematic phase. 

Fig.~4 displays representative normalized DLS correlation functions recorded in the nematic and twist-bend phases of $5~\mu$m thick samples of the LC mixture for geometries G1 and G2. In the ``splay" geometry (G1), a single overdamped fluctuation mode is detected in both N and $\mathrm{N_{TB}}$ phases. By scanning $\theta_s$, we determined $\Gamma_1^n \sim q^2$ with $\Gamma_1^n/q^2$ in the range $10^{-11} - 10^{-10}$~s$^{-1}$~m$^2$. Thus, splay fluctuations of the optic axis are hydrodynamic on both sides of the transition.

The spectrum and behavior of modes detected in geometry G2 are more interesting. In the nematic phase (above $T_{TB}$), two overdamped modes are observed in the range of $\theta_s$ studied: the expected hydrodynamic twist-bend director mode with relaxation rate $\Gamma_2^n \sim q^2$ (see measured $q^2$ dependence in Fig.~5) of order $\sim 10^3$~$s^{-1}$ and $\Gamma_2^n/q^2 \simeq 10^{-11}-10^{-10}$~s$^{-1}$~m$^2$, and a faster, nonhydrodynamic mode (Fig.~5) with $\Gamma_2^p \simeq 10^5$~s$^{-1}$ and independent of $q$. (The meaning of superscript $p$ will be clarified in the next section.) The relaxation rates of both modes were extracted from fits of the correlation data to double exponential decays.

\begin{figure}[tbp] 
\centering
\includegraphics[width=.8\columnwidth]{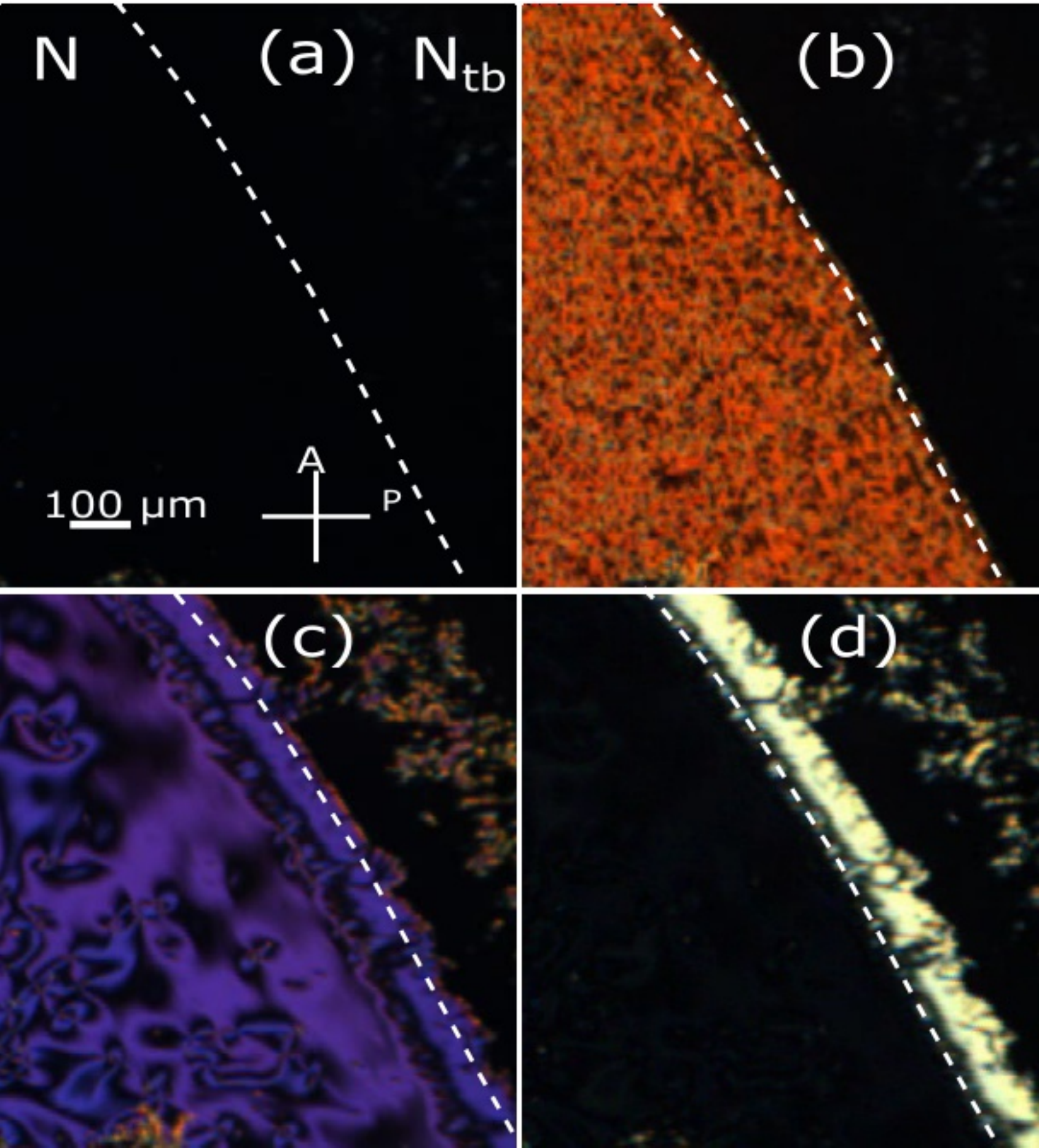}
\caption{(Color online) Polarizing microscope textures for a $5 \mu$m thick homeotropically aligned sample of the studied mixture. The optic axis is normal to the image plane, and the sample is placed between crossed polarizers. (a): Separate regions of nematic (N) and twist-bend ($\mathrm{N_{TB}}$) phases observed at the transition between the two; the boundary is marked by the dashed line. Both regions are uniform and dark, indicating high-quality homeotropic alignment of the director $\hat{\mathbf{n}}$ in the nematic and pitch axis $\hat{\mathbf{t}}$ in the $\mathrm{N_{TB}}$ phase. (b): Under an applied AC voltage (5 V @ 10 KHz), a second order Freedericsz transition (reorientation of $\hat{\mathbf{n}}$ in the center of the sample) is observed in the nematic region, while the $\mathrm{N_{TB}}$ region is unchanged. (c): Under higher voltage (7 V @ 10 KHz), the $\mathrm{N_{TB}}$ region undergoes a first order reorientation of $\hat{\mathbf{t}}$ in the form of nucleating toroidal focal conic domains (FCDs) and expanding stripes of splay and saddle splay deformations of $\hat{\mathbf{t}}$. (d): Several seconds after the voltage has been switched off, the nematic region relaxes back to the homeotropic state, whereas the $\mathrm{N_{TB}}$ region relaxes considerably slower.}
\end{figure}

\begin{figure}[tbp] 
\centering
\includegraphics[width=1.0\columnwidth]{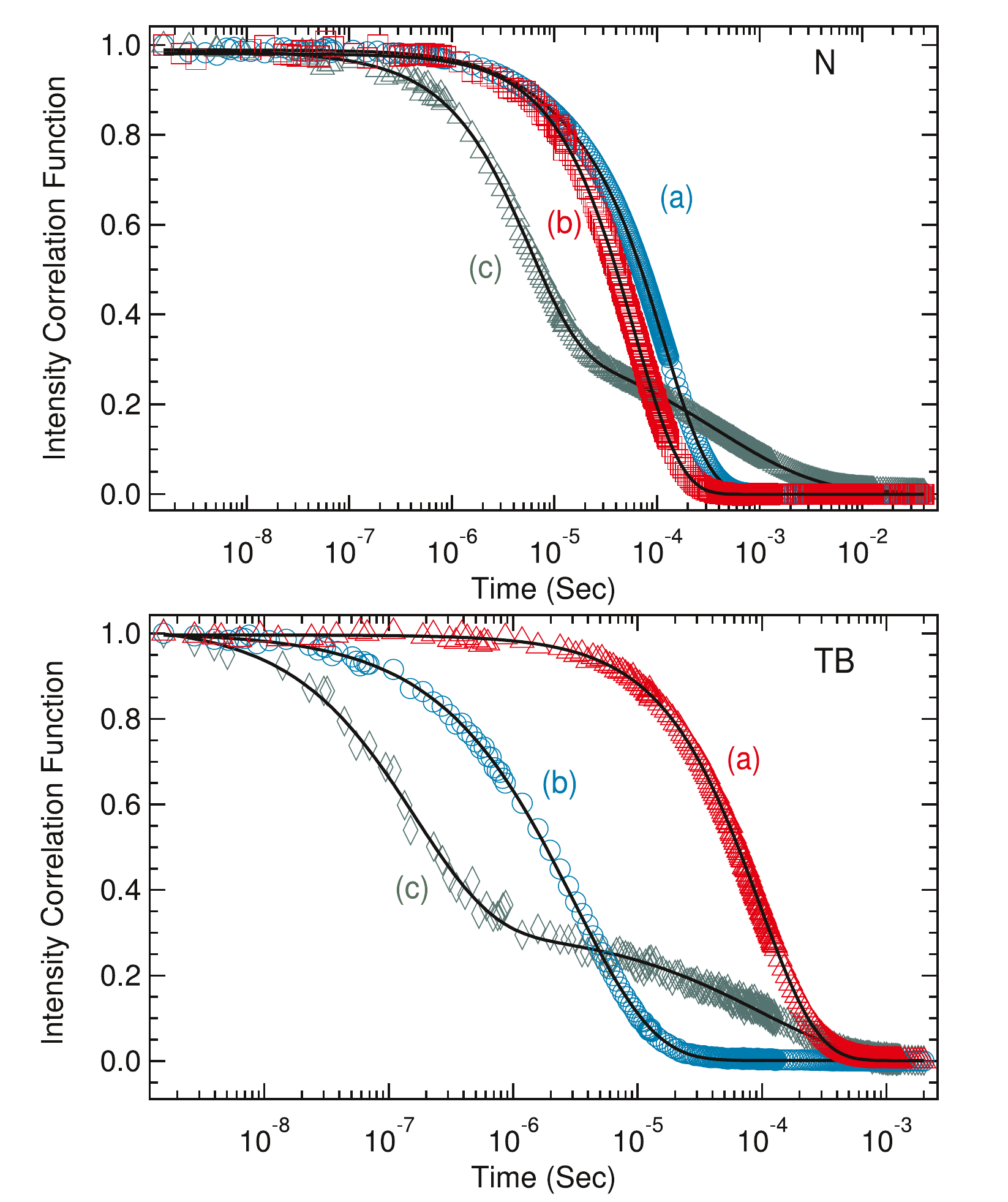}
\caption{(Color online) Top panel: Normalized homodyne DLS correlation functions taken in the nematic phase of the studied LC mixture for (a) geometry G2 with $T-T_{TB} = 16.2^\circ$C and angles $\theta_i = 15^\circ$, $\theta_s=40^\circ$, (b) G1 with $T-T_{TB} = 6.0^\circ$C  and $\theta_i = 0^\circ$, $\theta_s = 60^\circ$, and (c) G2 with $T-T_{TB} = 16.2^\circ$C and $\theta_i = 15^\circ$, $\theta_s=0^\circ$ (``dark" director geometry). Solid lines represent fits to a single exponential decay, except for (c), which is fit to a double exponential with the slower component stretched. Bottom panel: Normalized correlation data taken in the $\mathrm{N_{TB}}$ phase for (a) geometry G1 with $T-T_{TB} = -1.1^\circ$C and $\theta_i = 0^\circ$, $\theta_s = 60^\circ$, (b) G2 with $T-T_{TB} = -0.62^\circ$C and $\theta_i = 15^\circ$, $\theta_s=40^\circ$, and (c) G2 with $T-T_{TB} = -2.5^\circ$C and $\theta_i = 35^\circ$, $\theta_s = 0^\circ$ (``dark" director geometry). Solid lines are single exponential fits, except for a double exponential in (c).} 
\end{figure}

The presence of the fast mode in the DLS correlation function is most evident in the ``dark director" geometry where $\theta_s = 0^\circ$ (see data labeled (c) in top panel of Fig.~4), although it contributes weakly for $\theta_s \neq 0^\circ$. However, even in the ``dark" geometry where fluctuations in $\hat{\mathbf{n}}$ do not contribute to the DLS to first order, we still observe the decay of the slow director mode with a significant spread in its relaxation rate. (The fit in this case used a stretched exponential, with one additional fitting parameter.) Alignment mosaicity and a consequent broadening of the scattered wavevector $\mathbf{k}_s$ relative to $\hat{\mathbf{n}}$ could produce a ``leakage" of the slow director mode, but that does not account for the fact that no significant spread in $\Gamma_2^n \sim q^2$ is observed for $\theta_s$ off the ``dark" condition. An alternative scenario based on an intrinsic coupling between the fast and slow fluctuations is argued in the Discussion section below.

In the $\mathrm{N_{TB}}$ phase, the relaxation rates and $q$-dependence of the modes observed in geometry G2 change significantly. The twist-bend director mode, which dominates the scattering for $\theta_s \neq 0$, develops a large energy gap; its relaxation rate increases markedly below the transition ($T = T_{TB}$) to values in the $10^5 - 10^6$~s$^{-1}$ range, and, as evidenced in Fig.~5, becomes $q$-independent. Thus, below $T_{TB}$, the twist-bend mode crosses over from a hydrodynamic to nonhydrodynamic mode. As we shall demonstrate in the Discussion section, the magnitude of the gap is consistent with a modulation of $\hat{\mathbf{n}}$, whose period agrees with the FFTEM results \cite{Borshch_Nature} for the nanoscale periodic structure of the $\mathrm{N_{TB}}$ phase. Since the effective director (or optic axis) is the pitch axis $\hat{\mathbf{t}}$, for clarity we label its relaxation rate as $\Gamma_2^t$ (replacing $\Gamma_2^n$). 

\begin{figure}[tbp] 
\centering
\includegraphics[width=1.0\columnwidth]{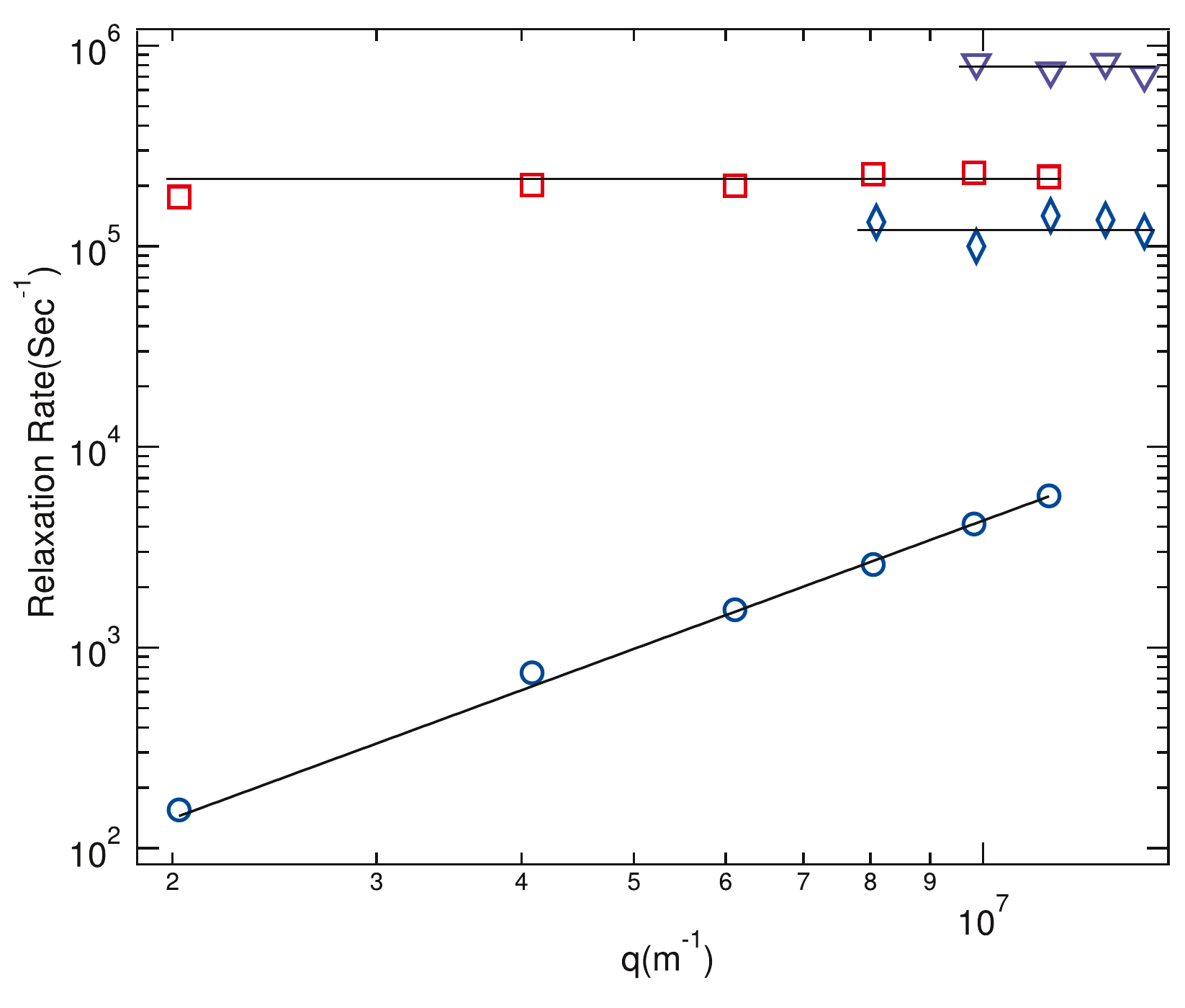}
\caption{(Color online) Dependence of the relaxation rates of the fluctuation modes detected in geometry G2 on the magnitude of the scattering vector $q$. Circles and squares correspond to relaxation rates $\Gamma_2^n$ and $\Gamma_2^p$ of the hydrodynamic director and nonhydrodynamic polarization modes detected in scattering geometry G2 in the middle of the nematic phase ($T-T_{TB} = 25^\circ$C). The slope of the line through the data on the log-log plot for $\Gamma_2^n$ is 2, indicating $\Gamma_2^n \sim q^2$. Diamonds and triangles correspond to relaxation rate $\Gamma_2^t$ of the nonhydrodynamic pitch axis fluctuations at temperatures $T-T_{TB} = -0.85^\circ$C and $-8.0^\circ$C, respectively, in the $\mathrm{N_{TB}}$ phase. These data are limited to higher $q$ (or $\theta_s$) due to a large component of background scattering at lower $q$, whose effect is exacerbated because of the low scattering intensity from fluctuations in the $\mathrm{N_{TB}}$ phase in the G2 geometry.}
\end{figure}

Correlation data taken in the ``dark director" geometry (G2 with $\theta_s = 0^\circ$) in the $\mathrm{N_{TB}}$ phase reveal a second, even faster nonhydrodynamic mode with a relaxation rate of $10^6-10^7$~s$^{-1}$ (see data labeled (c) in the bottom panel of Fig.~4), $\sim 10$ times higher than the values of $\Gamma_2^p$ for the fast mode in the nematic phase detected in the same geometry. Additionally, and again as in the nematic, a slow process -- with relaxation rate comparable to that of a hydrodynamic director mode -- also contributes to the correlation function. 

In both phases, the total scattering intensity in the ``dark" geometry, $\theta_s = 0^\circ$, is $\sim 10$ times weaker than the intensity for neighboring angles $\theta_s = \pm 10^\circ$, where the twist-bend director mode couples to the dielectric tensor and dominates the scattering.
 
\begin{figure}[tbp] 
\centering
\includegraphics[width=1.0\columnwidth]{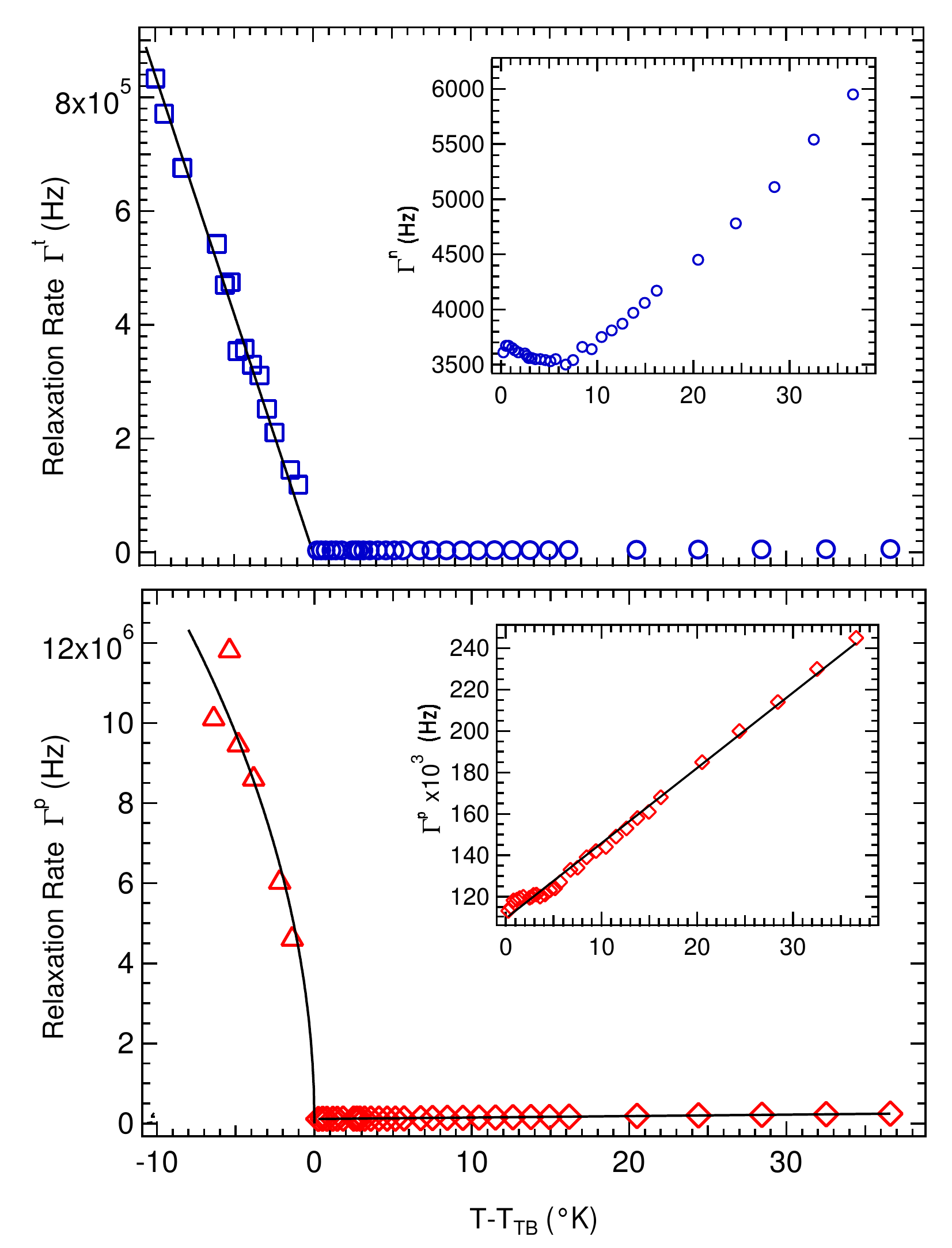}
\caption{(Color online) Top: Temperature dependence of relaxation rates associated with director fluctuations detected in scattering geometry G2 in the uniaxial nematic phase ($\Gamma^n$ for $T>T_{TB}$, circles in main figure and inset) and twist-bend phase ($\Gamma^t$ for $T<T_{TB}$, squares in main figure) with fixed $\theta_i$ and $\theta_s$. Bottom: Temperature dependence of the relaxation rate of polarization fluctuations in the nematic (diamonds in main figure and inset) and $\mathrm{N_{TB}}$ (triangles in main figure) phases. The solid lines in both panels are fits of $\Gamma^t$ and $\Gamma^p$ to calculated results from the coarse-grained free energy density model presented in the Theory section of the text.}
\end{figure}

Fig.~6 shows the temperature dependence of the relaxation rates for the two nonhydrodynamic modes ($\Gamma_2^t$ and $\Gamma_2^p$) in the $\mathrm{N_{TB}}$ phase, and for the nonhydrodynamic mode ($\Gamma_2^p$) and hydrodynamic director mode ($\Gamma_2^n$) in the nematic phase (see figure inset). These results were obtained from analysis of correlation data taken at fixed $\theta_i, \theta_s$ in geometry G2. The nonhydrodynamic modes clearly slow down significantly on approach to $T_{TB}$ from both sides of the transition, although on the low temperature side the present data are limited to temperatures futher than $1^\circ$C--$2^\circ$C from the transition.

\begin{figure}[tbp] 
\centering
\includegraphics[width=1.0\columnwidth]{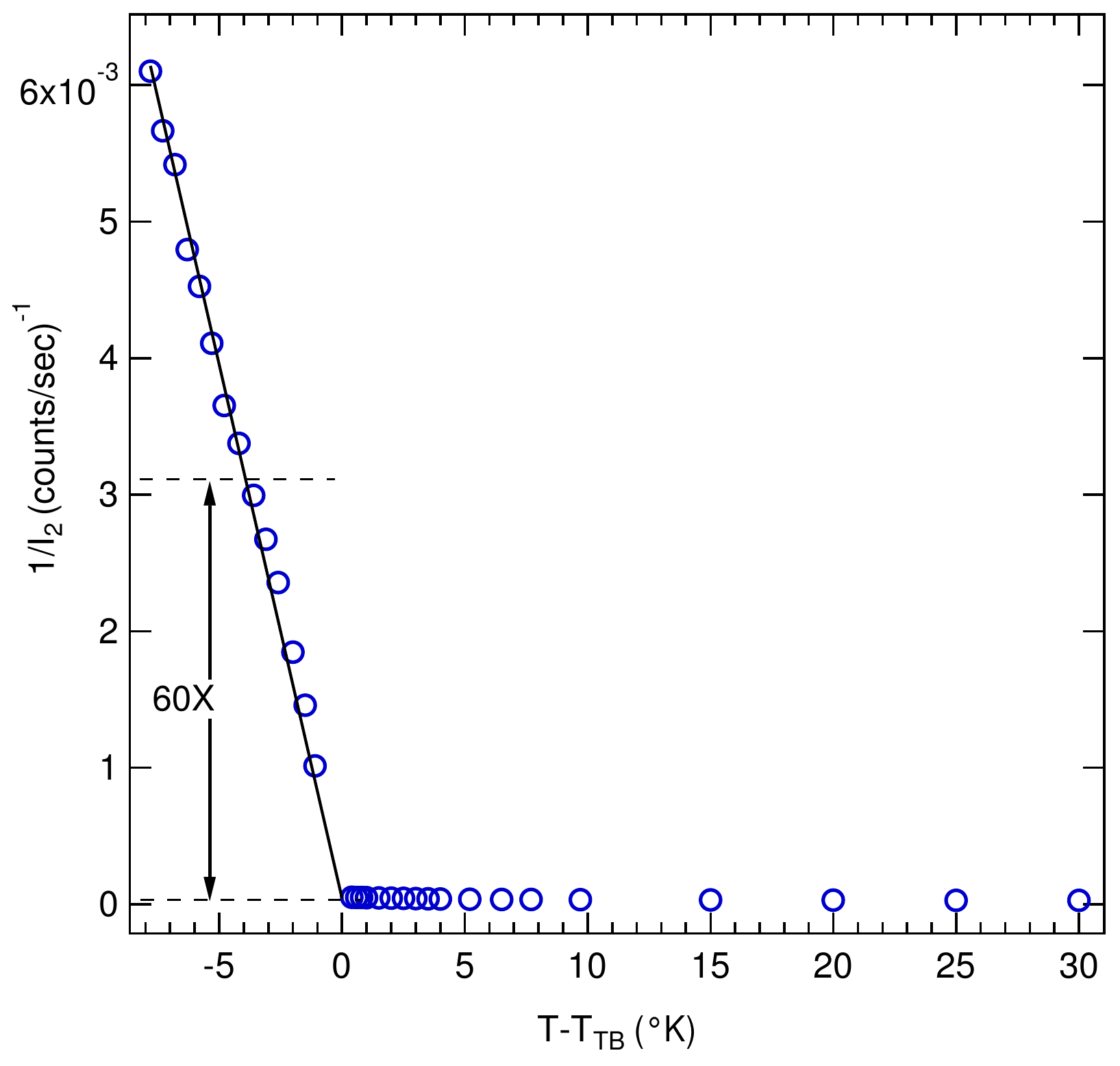}
\caption{(Color online) Temperature dependence of the inverse of the total scattering intensity $I_2^{-1}$ recorded in geometry G2 for $\theta_i=15^\circ, \theta_s = 40^\circ$. The solid line is linear fit of the data for $T<T_{TB}$.}
\end{figure}

Finally, the temperature dependence of the inverse total scattered intensity ($I_2^{-1}$), recorded in geometry G2, is plotted in Fig.~7. These data were taken at fixed $\theta_i=15^\circ$, $\theta_s = 40^\circ$, where the dominant signal in the $\mathrm{N_{TB}}$ phase comes from the nonhydrodynamic mode corresponding to $\Gamma_2^t$, and in the nematic phase from the hydrodynamic twst-bend director mode corresponding to $\Gamma_2^n$. As $T \rightarrow T_{TB}$ from below, the decrease in $I_2^{-1}$ mirrors the decrease in $\Gamma_2^t$ (Fig.~6).

\section{Theory}
A successful model for the experimental fluctuation spectrum must account for: (1) the crossover from two hydrodynamic and one nonhydrodynamic mode in the nematic to one hydrodynamic and two nonhydrodynamic modes in the $\mathrm{N_{TB}}$ phase; (2) the identity of the faster (nonhydrodynamic) mode detected in each phase; (3) the coupling of this fast process to slower director modes (evidenced in the data from the ``dark director" geometry); and (4) the temperature dependence of the relaxation rates of the nonhydrodynamic modes. To this end, we require a model free energy density for the nematic to twist-bend transition that contains relevant hydrodynamic and nonhydrodynamic fields, and the appropriate coupling between them.

Shamid et al \cite{Shamid_PRE} have recently analyzed the equilibrium behavior of such a model. The essential ingredient of their theory is a vector order parameter representing a polarization field $\mathbf{P}$ that originates, e.g., from the transverse dipole moment associated with the bent conformation of the dimer molecules that promotes the formation of the $\mathrm{N_{TB}}$ phase. It is convenient to use a dimensionless form for the order parameter, $\mathbf{p} = \mathbf{P}/P_{sat}$, where $P_{sat}$ corresponds to the saturated polarization at low temperature. 

The free energy density expanded in terms of the fields $\hat{\mathbf{n}}$ and $\mathbf{p}$ reads
\begin{eqnarray}
\label{fntb}
F_{NTB} &=& \frac{K_1}{2} (\nabla \cdot \hat{\mathbf{n}})^2 + \frac{K_2}{2} (\hat{\mathbf{n}} \cdot \nabla \times \hat{\mathbf{n}})^2\\
&&+ \frac{K_3}{2} [\hat{\mathbf{n}} \times (\nabla \times \hat{\mathbf{n}})]^2 + \frac{\mu}{2} |\mathbf{p}|^2 + \frac{\nu}{4} |\mathbf{p}|^4 \nonumber\\
&&+ \frac{\kappa}{2} (\nabla \mathbf{p})^2 - \Lambda [\hat{\mathbf{n}} \times (\nabla \times \hat{\mathbf{n}})] \cdot \mathbf{p} + \eta (\hat{\mathbf{n}} \cdot \mathbf{p})^2 . \nonumber
\end{eqnarray}
Here, $K_1$, $K_2$, and $K_3$ are the Frank elastic constants for splay, twist, and bend distortions of the director $\hat{\mathbf{n}}$.  The coefficient $\mu = \mu_0 (T - T_0)$ is the temperature-dependent Landau coefficient for the polarization $\mathbf{p}$ ($\mu_0$ being a constant), while $\nu>0$ is a higher-order, temperature-independent Landau coefficient.  The elastic constant $\kappa$ penalizes spatial distortions in $\mathbf{p}$, and the coefficient $\Lambda$ couples $\mathbf{p}$ with bend distortions.  The last term (not included in Ref.~\cite{Shamid_PRE}), with $\eta > 0$, favors polarization perpendicular to the nematic director and is consistent with bend flexoelectricity. Because $\mathbf{p}$ is defined to be dimensionless, the Landau coefficients $\mu$ and $\nu$ carry the same units, and $\kappa$ has the same units as the Frank constants.

In the $\mathrm{N_{TB}}$ phase, the director field has the heliconical modulation
\begin{equation}
\hat{\mathbf{n}} = \hat{\mathbf{z}}\cos\beta+\hat{\mathbf{x}}\sin\beta\cos(q_0 z)+\hat{\mathbf{y}}\sin\beta\sin(q_0 z),
\label{ntbgroundstate}
\end{equation}
with pitch wavenumber $q_0$ and cone angle $\beta$.  (Note that $\sin\beta$ was called $a$ in Ref. \cite{Shamid_PRE}.)  Likewise, the polarization field has the helical modulation
\begin{equation}
\mathbf{p} =\hat{\mathbf{x}}p_0\sin(q_0 z)-\hat{\mathbf{y}}p_0\cos(q_0 z),
\end{equation}
with magnitude $p_0$, perpendicular to $\hat{\mathbf{n}}$ and to the pitch axis $\hat{\mathbf{z}}$, as shown in Fig. 1 (left side).  In the nematic phase, $\beta$ and $p_0$ are both zero while $q_0$ is undefined; in the $\mathrm{N_{TB}}$ phase, these quantities all become non-zero.

To find the ground state, we must insert Eqs.~(2) and (3) into Eq.~(1) for $F_{NTB}$ and then minimize with respect to $q_0$, $\beta$, and $p_0$.  For this calculation, we repeat the work of Ref. \cite{Shamid_PRE} and generalize it to the case of weak polar elastic constant $\kappa$, which will turn out to be physically relevant.  First, minimization with respect to $q_0$ gives 
\begin{equation}
q_0 = \frac{\Lambda p_0 \sin\beta \cos\beta}{\kappa p_0^2 + K_3 \sin^2\beta \cos^2\beta + K_2 \sin^4 \beta},
\label{q0min}
\end{equation}
and minimization with respect to $\beta$ gives
\begin{equation}
\sin^2 \beta = -\frac{\kappa p_0^2}{K_2}+\sqrt{\frac{\kappa p_0^2}{K_2}\left(1+\frac{\kappa p_0^2}{K_2}\right)}.
\label{betamin}
\end{equation}
Equation~(\ref{betamin}) can be compared with the experiment of Ref.~\cite{Borshch_Nature}, which shows the cone angle $\beta \lsim 10^\circ$ within the temperature range covered by our DLS data. This result implies that $p_0 (\kappa/K_2)^{1/2} \lsim 0.03$.  Because $p_0$ is a scaled polarization, which grows to order 1 at low temperature, we estimate that $(\kappa/K_2)^{1/2} \simeq 0.03$, which shows that the polarization elasticity is small compared with the Frank director elasticity.

Substituting Eqs.~(\ref{q0min}) and (\ref{betamin}) into the free energy density and expanding for small $p_0$ and $\kappa$ gives
\begin{eqnarray}
F_{NTB} &=& \frac{1}{2}\left[\mu_0 (T-T_0)-\frac{\Lambda^2}{K_3}\right]p_0^2 \nonumber\\
&& + \frac{\Lambda^2 \kappa^{1/2} K_2^{1/2}}{K_3^2}|p_0|^3+\frac{1}{4}\nu p_0^4.
\label{Feffective}
\end{eqnarray}
From this form of the effective free energy density, we can see that there is a second-order transition from the nematic to the $\mathrm{N_{TB}}$ phase at the temperature
\begin{equation}
T_{TB} = T_0 + \frac{\Lambda^2}{K_3 \mu_0}.
\label{TTB}
\end{equation}

This transition is unusual because the relative magnitudes of the cubic and quartic terms in Eq.~(\ref{Feffective}) depends on the relative smallness of $p_0$ and $\kappa$.  Close to the transition, where $p_0\ll(\Lambda^2 \kappa^{1/2} K_2^{1/2})/(K_3^2 \nu)$, the cubic term dominates over the quartic term.  By minimizing the effective free energy, we see that $p_0$ depends on temperature as
\begin{equation}
p_0(T) = \frac{K_3^2 \mu_0 (T_{TB}-T)}{3\Lambda^2 \kappa^{1/2} K_2^{1/2}}.
\label{p0scaling1}
\end{equation}
This result is consistent with the scaling reported in Ref.~\cite{Shamid_PRE}, with a slight correction in the numerical coefficient.  By contrast, farther from the transition, where $p_0\gg(\Lambda^2 \kappa^{1/2} K_2^{1/2})/(K_3^2 \nu)$, the quartic term dominates over the cubic term, and the prediction for $p_0$ becomes
\begin{equation}
p_0(T) = \sqrt{\frac{\mu_0 (T_{TB}-T)}{\nu}}.
\label{p0scaling2}
\end{equation}
From the general form for $p_0$,
\begin{equation}
p_0 (T) = - \frac{3 \Lambda^2 (\kappa K_2)^{1/2}}{2 K_3^2 \nu} + \sqrt{ \frac{9 \Lambda^4 \kappa K_2}{4 K_3^4 \nu^2} + \frac{\mu_0}{\nu} (T_{TB} - T)}, \nonumber
\end{equation}
the crossover between these two regimes occurs at
\begin{equation}
(T_{TB}-T)=\frac{9\Lambda^4 \kappa K_2}{4 K_3^4 \mu_0 \nu}.
\end{equation}
We will see below that the crossover point is extremely close to the transition, so that all of the experimental data are taken in the regime governed by Eq.~(\ref{p0scaling2}) rather than Eq.~(\ref{p0scaling1}). 

As an aside, this theory can easily be modified to describe a first-order transition between the nematic and $\mathrm{N_{TB}}$ phases, by changing the fourth-order coefficient $\nu$ to a negative value and adding a sixth-order term to $F_{NTB}$ in Eq.~(\ref{fntb}).  We have not done so here, because the DLS data give no indication of a first-order transition.  However, such a modification might be useful for analyzing the nematic-$\mathrm{N_{TB}}$ transition in other systems.

Now that we have determined the ground state, we will consider fluctuations about the ground state in the nematic and $\mathrm{N_{TB}}$ phases.

\subsection{Nematic phase}

In the nematic phase, we must consider fluctuations in the director field about the ground state $\hat{\mathbf{n}} = \hat{\mathbf{z}}$, and fluctuations in the polarization about the ground state $\mathbf{p} = 0$.  At lowest order, these fluctuations can be described by $\mathbf{\delta n}(\mathbf{r}) = (n_x, n_y, 0)$ and $\mathbf{\delta p}(\mathbf{r}) = (p_x, p_y, p_z)$.  We insert these expressions into the free energy $F_{NTB}$ (Eq.~(1)), and expand to quadratic order in the fluctuating components.  We then Fourier transform from position $\mathbf{r}$ to wavevector $\mathbf{q}$, and express the free energy as a quadratic form in $n_x(\mathbf{q})$, $n_y(\mathbf{q})$, $p_x(\mathbf{q})$, $p_y(\mathbf{q})$, and $p_z(\mathbf{q})$,
\begin{widetext}
\begin{equation}
F = \frac{1}{2} \sum_{\mathbf{q}}
\begin{pmatrix}
 n_{x\mathbf{q}}\\
 p_{x\mathbf{q}} \\
 n_{y\mathbf{q}}\\
 p_{y\mathbf{q}}\\ 
 p_{z\mathbf{q}}
\end{pmatrix}^\dagger
\begin{pmatrix}
 K_1 q_x^2 + K_2 q_y^2 + K_3 q_z^2 & -i \Lambda q_z & 0& 0& 0\\ 
 i \Lambda q_z & \mu + \kappa |\mathbf{q}|^2 & 0& 0& 0\\ 
 0 & 0 & K_2 q_x^2 + K_1 q_y^2 + K_3 q_z^2 & -i \Lambda q_z& 0\\  
 0 &0 &i \Lambda q_z & \mu + \kappa |\mathbf{q}|^2 & 0\\    
 0 &0 &0 &0 & 2\eta + \mu + \kappa |\mathbf{q}|^2
\end{pmatrix}
\begin{pmatrix}
 n_{x\mathbf{q}}\\
 p_{x\mathbf{q}} \\
 n_{y\mathbf{q}}\\
 p_{y\mathbf{q}}\\ 
 p_{z\mathbf{q}}
\end{pmatrix}.
\end{equation}  
\end{widetext}
By diagonalizing this quadratic form, we obtain five normal modes:

(1) One hydrodynamic mode is primarily splay-bend director fluctuations, combined with some polarization fluctuations.  Its relaxation rate is the ratio of the free energy eigenvalue to the relevant viscosity coefficient $\gamma_n$, which gives
\begin{equation}
\Gamma^n_1=\frac{K_1 q_\perp^2 + K_3^\text{eff} q_z^2}{\gamma_n}
\end{equation}
in the limit of long wavelength (small $\mathbf{q}$).  Here,
\begin{equation}
K_3^\text{eff}=K_3-\frac{\Lambda^2}{\mu}=K_3-\frac{\Lambda^2}{\mu_0 (T - T_0)}
\label{K3eff}
\end{equation}
is the renormalized bend elastic constant \cite{Shamid_PRE}, which shows the effect of coupling the director to the polarization.  This effect accounts for the softening of bend fluctuations observed in earlier DLS studies of the director modes when $T \rightarrow T_{TB}$ from the nematic side \cite{hyd_mode}. Specifically, Eqs.~(\ref{K3eff}) and (\ref{TTB}) imply $K_3^\text{eff} = 0$ at $T = T_{TB}$. 

(2) Another hydrodynamic mode is primarily twist-bend director fluctuations, combined with some polarization fluctuations.  Its relaxation rate is 
\begin{equation}
\Gamma^n_2=\frac{K_2 q_\perp^2 + K_3^\text{eff} q_z^2}{\gamma_n},
\end{equation}
again with the renormalized bend elastic constant $K_3^\text{eff}$.

(3, 4) Two nonhydrodynamic modes are mostly polarization fluctuations $p_x$ and $p_y$, combined with some director fluctuations.  In the limit of $\mathbf{q}\to0$, these modes have relaxation rate
\begin{equation}
\Gamma^p = \frac{\mu}{\gamma_p} = \frac{\mu_0 (T - T_0)}{\gamma_p}.
\label{Gammapinnematic}
\end{equation}

(5) Another nonhydrodynamic mode is polarization $p_z$ by itself.  In the limit of $\mathbf{q}\to0$, it has relaxation rate
\begin{equation}
\Gamma^{p^\prime} = \frac{2 \eta + \mu}{\gamma_{p^\prime}} = \frac{2 \eta + \mu_0 (T - T_0)}{\gamma_{p^\prime}}.
\label{Gammapprimeinnematic}
\end{equation}
Here, $\gamma_p$ and $\gamma_{p^\prime}$ are the mode viscosities.

Overall, we should emphasize the contrast between the nematic phase of the $\mathrm{N_{TB}}$-forming material studied here and a typical nematic phase.  In the $\mathrm{N_{TB}}$-forming material, we observe a nonhydrodynamic mode with a relaxation rate that decreases with temperature, as the system approaches the transition to the $\mathrm{N_{TB}}$ phase.  The theory attributes this mode to polarization fluctuations, which become less energetically costly as the system develops incipient polar order.  By contrast, in a typical nematic phase, no such mode can be observed in DLS experiments; presumably polarization fluctuations decay too rapidly to be detected.

\subsection{Twist-bend phase}

In the $\mathrm{N_{TB}}$ phase, the analysis of normal modes is complicated because of the nonuniform, modulated director structure. However, as mentioned in the Introduction, we can simplify this calculation through a \emph{coarse-graining} approximation, which averages over the director modulation to find the larger-scale properties of the phase. Such coarse graining has previously been done for the cholesteric phase \cite{Lubensky}, and it shows that the cholesteric has the same macroscopic elastic properties as a smectic phase. In this section, we generalize the coarse-graining procedure to the more complex case of the $\mathrm{N_{TB}}$ phase. Indeed, it should be an even better approximation for the $\mathrm{N_{TB}}$ than for the cholesteric phase, because the pitch of the $\mathrm{N_{TB}}$ is so short.

The basic concept of the coarse-graining procedure is illustrated in Fig.~1.  We suppose that the director field has a rapid heliconical modulation with respect to a local orthonormal reference frame $(\hat{\mathbf{e}}_1(\mathbf{r}),\hat{\mathbf{e}}_2(\mathbf{r}),\hat{\mathbf{t}}(\mathbf{r}))$, and this orthonormal frame varies slowly in space.  Furthermore, the heliconical modulation might be displaced upward or downward by a phase $\phi(\mathbf{r})$, which also varies slowly in space.  Hence, the director field can be written as
\begin{eqnarray}
\label{directoransatz}
\hat{\mathbf{n}}(\mathbf{r}) &=& \hat{\mathbf{t}}(\mathbf{r})\cos\beta + \hat{\mathbf{e}}_1(\mathbf{r})\sin\beta\cos(q_0 z+\phi(\mathbf{r}))\nonumber\\
&& + \hat{\mathbf{e}}_2(\mathbf{r})\sin\beta\sin(q_0 z+\phi(\mathbf{r})).
\end{eqnarray}
In this expression, $\hat{\mathbf{t}}(\mathbf{r})$ is the coarse-grained director, which would be measured in any experiment that averages over the nanoscale heliconical modulation.  By analogy with the director field, the polarization field has a rapid helical modulation with respect to the same local orthonormal reference frame, which can be written as
\begin{eqnarray}
\label{polarizationansatz}
\mathbf{p}(\mathbf{r}) &=& \hat{\mathbf{e}}_1(\mathbf{r})p_0\sin(q_0 z+\phi(\mathbf{r}))
\nonumber\\
&&- \hat{\mathbf{e}}_2(\mathbf{r})p_0\cos(q_0 z+\phi(\mathbf{r}))+\mathbf{\delta p}(\mathbf{r}).
\end{eqnarray}
Here, $\mathbf{\delta p}(\mathbf{r})=\delta p_x \hat{\mathbf{x}}+\delta p_y \hat{\mathbf{y}}+\delta p_z \hat{\mathbf{z}}$ is a fluctuating additional contribution to the polarization, which varies slowly in space.  It is allowed because $\mathbf{p}$ is not restricted to be a unit vector.  The contribution $\mathbf{\delta p}(\mathbf{r})$ is the coarse-grained polarization, which would be measured in any experiment that averages over the nanoscale helical modulation.

From Eqs.~(\ref{directoransatz}--\ref{polarizationansatz}), we can see that the pseudo-layers are surfaces of constant $q_0 z+\phi(\mathbf{r})=q_0(z-u(\mathbf{r}))$, where $u(\mathbf{r})=-\phi(\mathbf{r})/q_0$ is the local pseudo-layer displacement.  The local helical axis (or pseudo-layer normal) is given by the gradient
\begin{equation}
\hat{\mathbf{N}}(\mathbf{r})=\frac{\mathbf{\nabla}(q_0 z+\phi(\mathbf{r}))}{|\mathbf{\nabla}(q_0 z+\phi(\mathbf{r}))|}
=\frac{\hat{\mathbf{z}}-\mathbf{\nabla}u}{|\hat{\mathbf{z}}-\mathbf{\nabla}u|}.
\end{equation}

We now consider the case of a well-aligned sample, as in a light-scattering experiment.  In this case, the coarse-grained director $\hat{\mathbf{t}}(\mathbf{r})$ has small fluctuations about $\hat{\mathbf{z}}$, while the phase $\phi(\mathbf{r})$ and coarse-grained polarization $\mathbf{\delta p}(\mathbf{r})$ have small fluctuations around $0$.  The full orthonormal reference frame can be written as
\begin{align}
& \hat{\mathbf{e}}_1(\mathbf{r}) = \left(1-\textstyle{\frac{1}{2}}t_x^2\right)\hat{\mathbf{x}}-\textstyle{\frac{1}{2}}t_x t_y\hat{\mathbf{y}}-t_x\hat{\mathbf{z}},\nonumber\\
\label{orthonormalbasis}
& \hat{\mathbf{e}}_2(\mathbf{r}) = -\textstyle{\frac{1}{2}}t_x t_y\hat{\mathbf{x}}+\left(1-\textstyle{\frac{1}{2}}t_y^2\right)\hat{\mathbf{y}}-t_y\hat{\mathbf{z}},\\
& \hat{\mathbf{t}}(\mathbf{r}) = t_x\hat{\mathbf{x}}+t_y\hat{\mathbf{y}}+\left(1-\textstyle{\frac{1}{2}}t_x^2-\textstyle{\frac{1}{2}}t_y^2\right)\hat{\mathbf{z}},\nonumber
\end{align}
to quadratic order in $t_x(\mathbf{r})$ and $t_y(\mathbf{r})$.  One might think that another variable would be needed to specify the vectors $\hat{\mathbf{e}}_1$ and $\hat{\mathbf{e}}_2$ in the plane perpendicular to $\hat{\mathbf{t}}$.  However, rotations in this plane can be included in the choice of the phase $\phi$.  As discussed in Ref.~\cite{Lubensky} for the cholesteric case, such rotations are analogous to gauge transformations.  Hence, we make the specific choice of gauge in Eq.~(\ref{orthonormalbasis}).  With this choice, our orthonormal basis has small fluctuations away from $(\hat{\mathbf{x}},\hat{\mathbf{y}},\hat{\mathbf{z}})$.

We insert Eqs.~(\ref{directoransatz}--\ref{polarizationansatz}) for the director and polarization fields, together with Eq.~(\ref{orthonormalbasis}) for the orthonormal basis, into Eq.~(\ref{fntb}) for the free energy of the $\mathrm{N_{TB}}$ phase.  We then make the coarse-graining approximation:  We integrate over the rapid variations of $\cos q_0 z$ and $\sin q_0 z$, assuming that the slowly varying fields are constant over the length scale of the pitch.  We thus obtain an effective free energy in terms of the six coarse-grained variables $\phi(\mathbf{r})$, $t_x(\mathbf{r})$, $t_y(\mathbf{r})$, $\delta p_x(\mathbf{r})$, $\delta p_y(\mathbf{r})$, and $\delta p_z(\mathbf{r})$.  We expand the free energy to quadratic order in these fields, and Fourier transform it from position $\mathbf{r}$ to wavevector $\mathbf{q}$, to obtain
\begin{equation}
F = \frac{1}{2} \sum_{\mathbf{q}}
\begin{pmatrix}
 \phi_{\mathbf{q}} \\
 t_{x\mathbf{q}}\\
 \delta p_{y\mathbf{q}} \\
 t_{y\mathbf{q}}\\
 \delta p_{x\mathbf{q}}\\ 
 \delta p_{z\mathbf{q}}
\end{pmatrix}^\dagger
\mathbf{M}(\mathbf{q})
\begin{pmatrix}
 \phi_{\mathbf{q}} \\
 t_{x\mathbf{q}}\\
 \delta p_{y\mathbf{q}} \\
 t_{y\mathbf{q}}\\
 \delta p_{x\mathbf{q}}\\ 
 \delta p_{z\mathbf{q}}
\end{pmatrix}.
\label{fcg}
\end{equation}  
Here, $\mathbf{M}(\mathbf{q})$ is a matrix of wavevector-dependent coefficients, which must be diagonalized to find the normal modes.

It is most convenient to understand the mode structure in the limit of $\mathbf{q}\to0$.  In this limit, the matrix simplifies to the block-diagonal form
\begin{equation}
\bm{M}(0) =
\begin{pmatrix}
 0       &  0 &  0&  0&  0&  0 \\
 0 & m_{22}  & m_{23} & 0& 0& 0\\ 
 0 & m_{32} & m_{33} & 0& 0& 0\\ 
 0 &0 & 0 & m_{44} & m_{45}& 0\\  
  0 &0 &0 &m_{54} & m_{55} & 0\\    
  0 &0 &0 &0 &0 &  m_{66}
\end{pmatrix},
\label{m0_matrix}
\end{equation}  
where
\begin{align}
& m_{22} = m_{44} = p_0 q_0 \Lambda \sin\beta + \textstyle{\frac{1}{2}}(K_1 + K_2 - 2 K_3) q_0^2 \sin^2\beta, \nonumber\\
& m_{33} = m_{55} = \mu + 2 \nu p_0^2 + \eta \sin^2\beta, \nonumber\\
& m_{23} = m_{32} = -m_{45} = -m_{54} = - \textstyle{\frac{1}{2}} q_0 \Lambda \sin^2\beta, \nonumber\\
& m_{66} =  (2 \eta + \mu) + \nu p_0^2 - 2 \eta \sin^2\beta.
\label{matrixelements}
\end{align}
From this block-diagonal form, we can extract the following six normal modes:

\begin{figure*}[tbp] 
\includegraphics[width=.66\columnwidth]{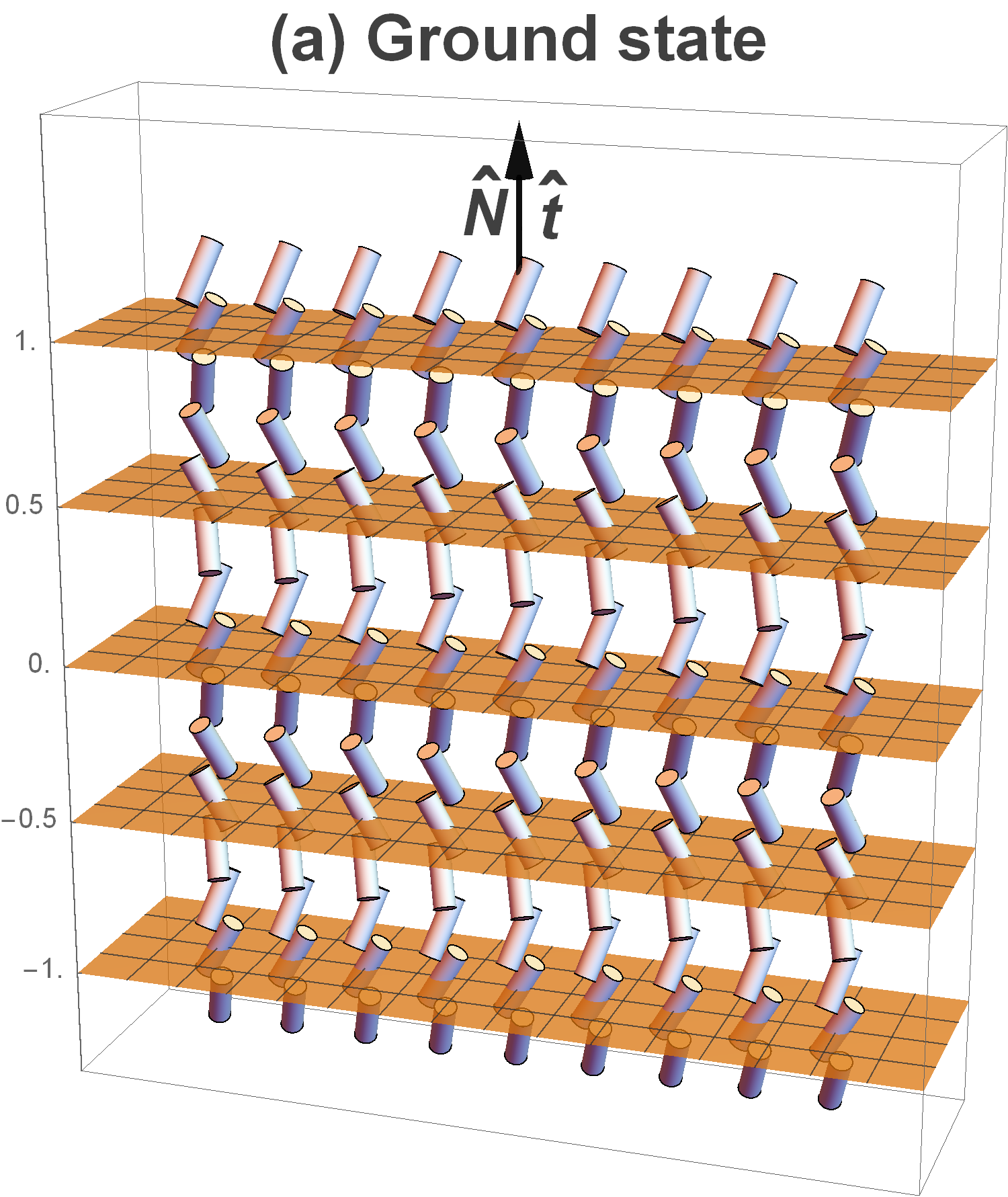}
\includegraphics[width=.66\columnwidth]{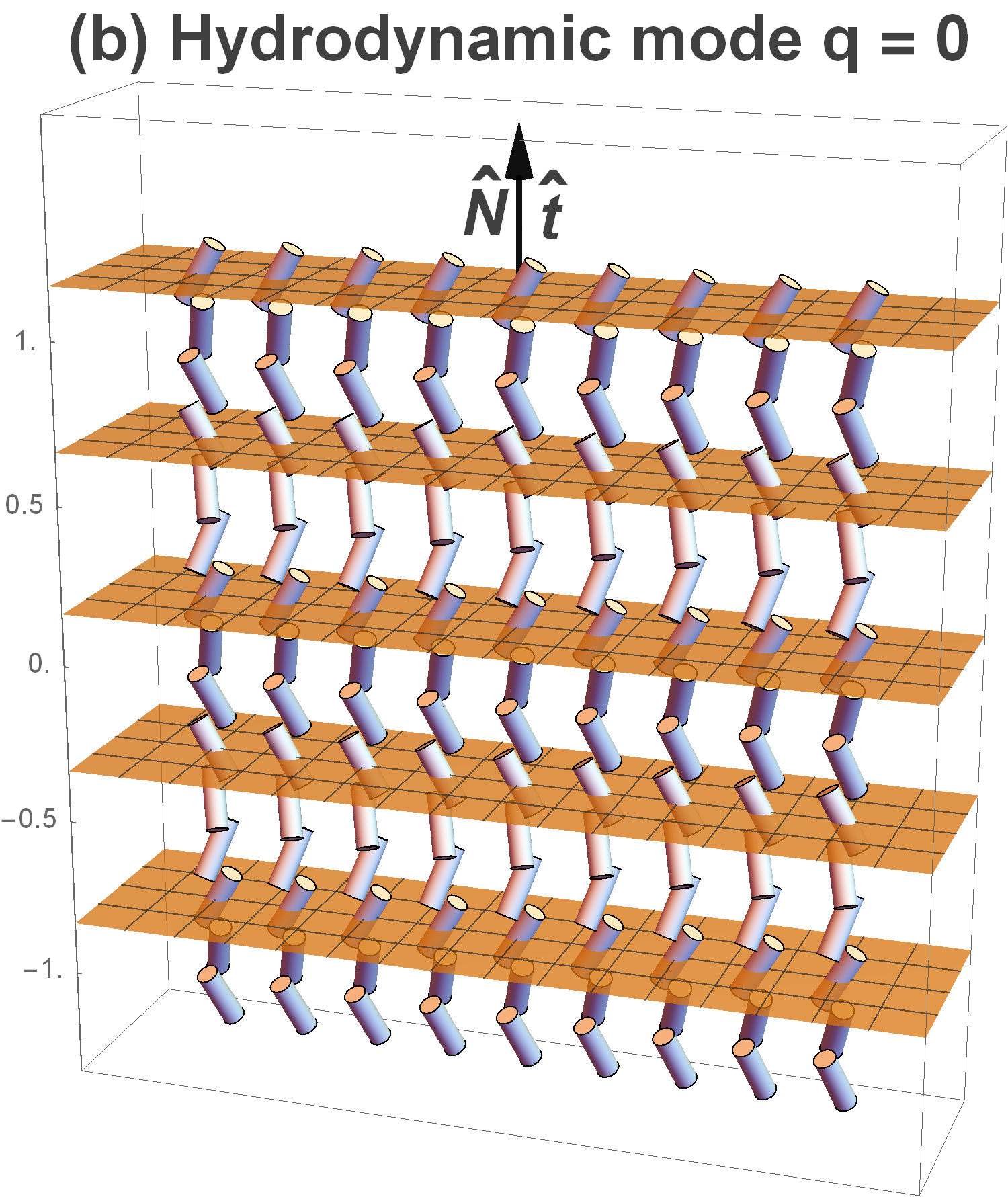}
\includegraphics[width=.66\columnwidth]{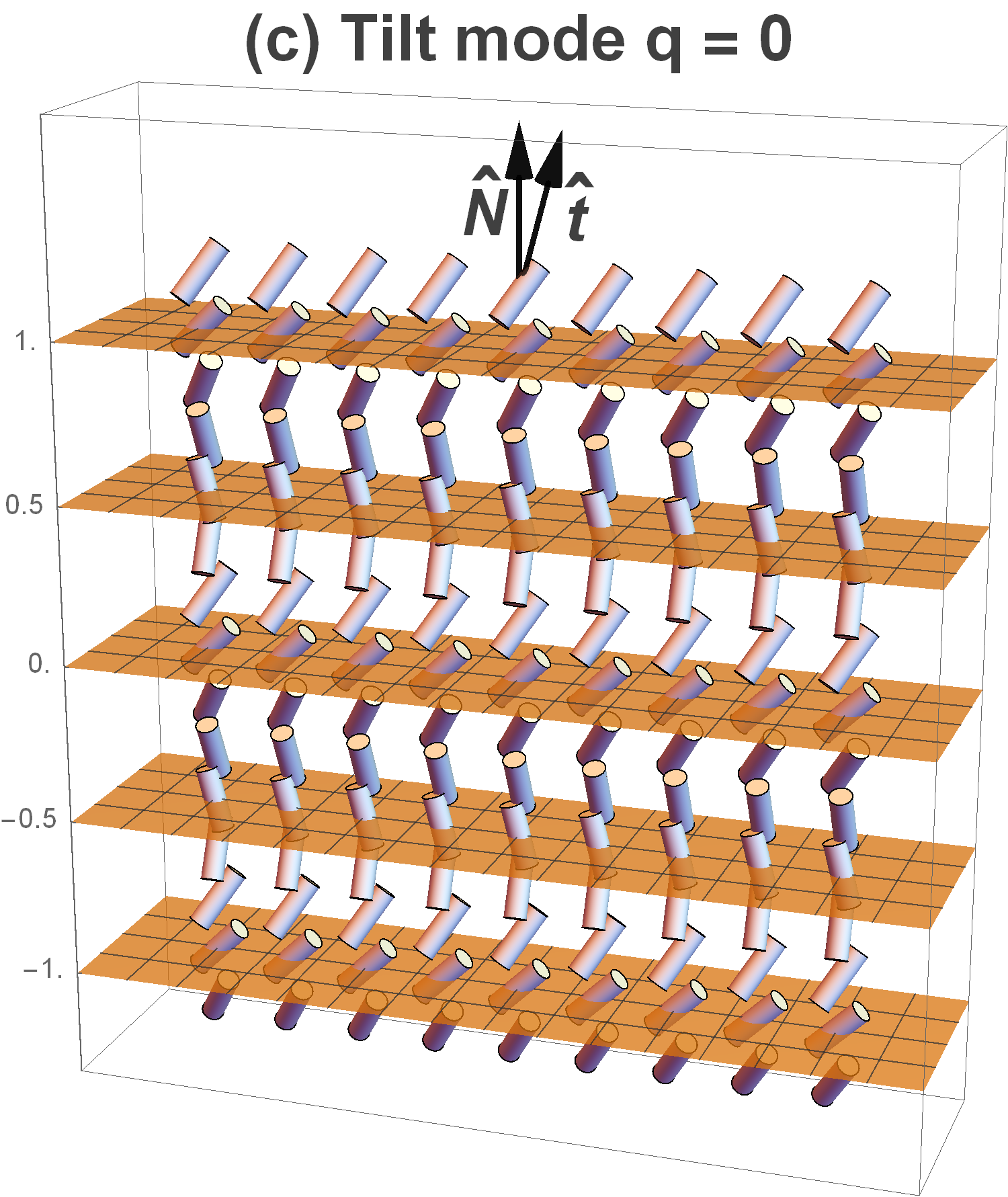}
\includegraphics[width=.66\columnwidth]{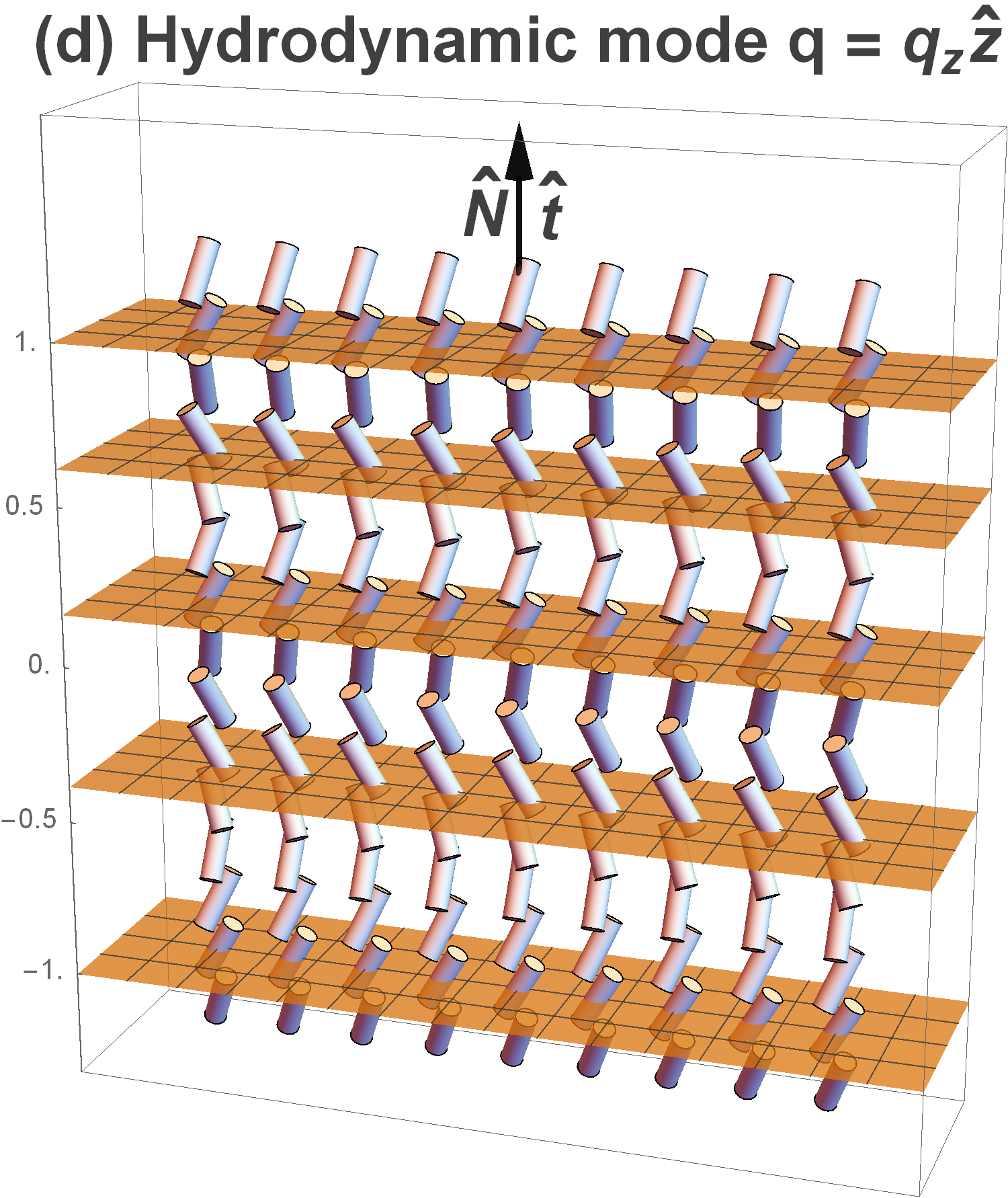}
\includegraphics[width=.66\columnwidth]{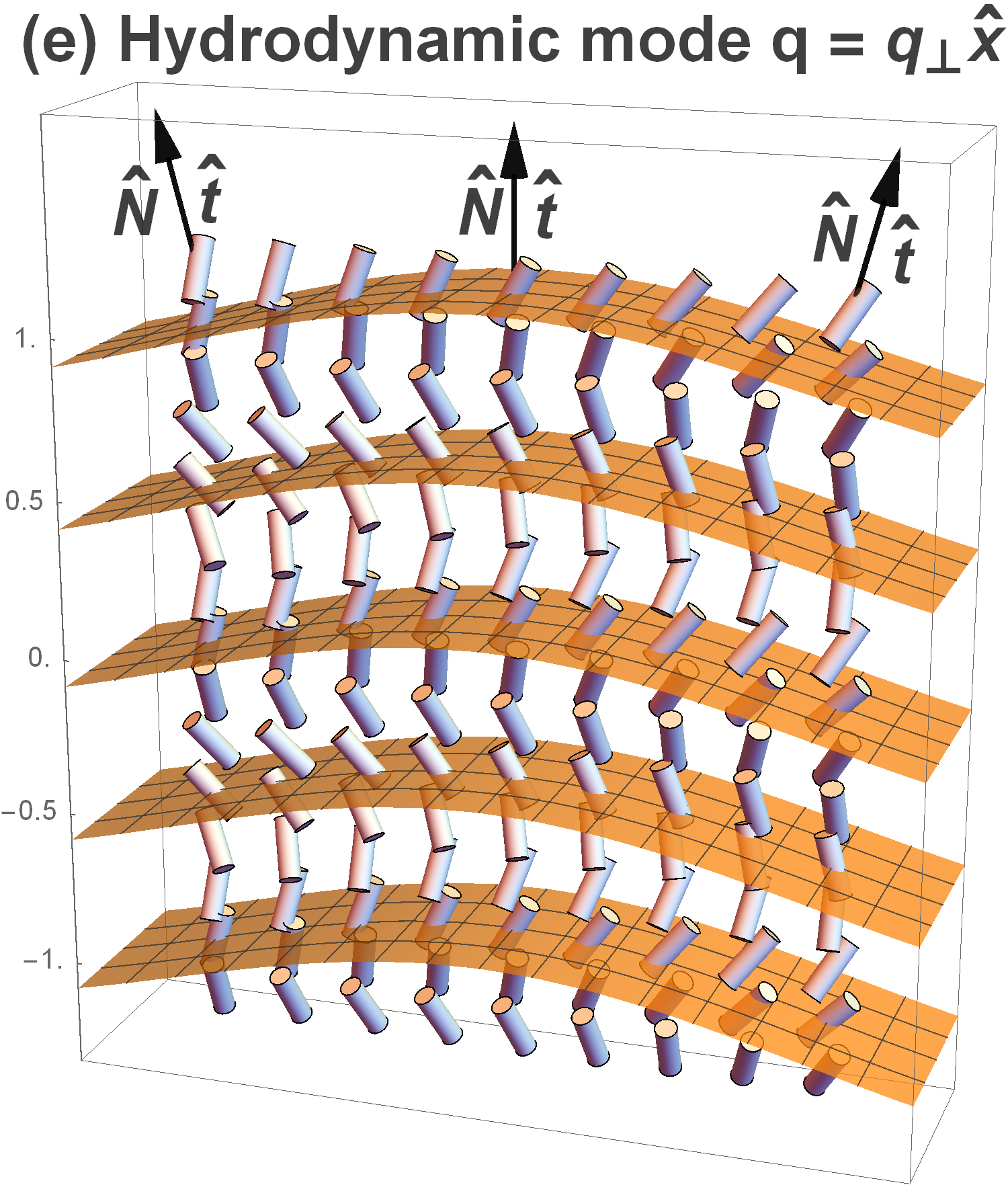}
\caption{(Color online) Visualization of fluctuating modes in the $\mathrm{N_{TB}}$ phase.  Small cylinders represent the heliconical director field $\hat{\mathbf{n}}(\mathbf{r})$, and surfaces represent the pseudo-layers.  (a) Ground state.  (b) Hydrodynamic mode with wavevector $\mathbf{q}=0$, with uniform rotation of $\hat{\mathbf{n}}(\mathbf{r})$ and hence uniform displacement of pseudo-layers; this mode has no energy cost with respect to the ground state.  (c) Nonhydrodynamic tilt mode, with the coarse-grained director $\hat{\mathbf{t}}$ (average of $\hat{\mathbf{n}}(\mathbf{r})$) tilted with respect to pseudo-layer normal.  (d) Hydrodynamic mode with $\mathbf{q}=q\hat{\mathbf{z}}$, with $z$-dependent rotation of $\hat{\mathbf{n}}(\mathbf{r})$ and $z$-dependent displacement of pseudo-layers (leading to compression and dilation).  (e) Hydrodynamic mode with $\mathbf{q}=q\hat{\mathbf{x}}$, with $x$-dependent rotation of $\hat{\mathbf{n}}(\mathbf{r})$ and $x$-dependent displacement of pseudo-layers (leading to curvature), accompanied by tilt so that $\hat{\mathbf{t}}$ remains normal to pseudo-layers.}
\end{figure*}

(1) The phase $\phi=-u/q_0$ is itself a normal mode.  This mode is hydrodynamic, with zero energy (and zero relaxation rate) in the limit of $\mathbf{q}\to0$.  It is analogous to the layer displacement of a smectic-A phase, which costs zero energy for uniform displacement.  It is also analogous to the hydrodynamic director mode in a cholesteric phase (which is called the pure twist mode in the theory of cholesteric light scattering \cite{deGennes}).  It is visualized in terms of pseudo-layers in Fig. 8a,b.

(2, 3) The coarse-grained director tilt $t_x$ and polarization $\delta p_y$ are coupled by the helicity of the $\mathrm{N_{TB}}$ phase.  Together, they form a pair of normal modes, both of which are non-hydrodynamic, with non-zero energy (and non-zero relaxation rate) in the limit of $\mathbf{q}\to0$. In the limit of weak coupling, which is given by the criterion $m_{22} m_{33} \gg m_{23}^2$,
\begin{subequations}
\label{GammatGammapintwistbend}
\begin{eqnarray}
\Gamma^t &=& m_{22} / \gamma_t,\\
\Gamma^p &=& m_{33} / \gamma_p.
\end{eqnarray}
\end{subequations}
Here, $\gamma_t$ and $\gamma_p$ are phenomenological viscosities associated with the normal modes.  The two modes are analogous to tilt and polarization fluctuations in a chiral smectic-A phase.  The tilt mode is also analogous to the non-hydrodynamic director mode in a cholesteric phase (which is called the umbrella mode in the theory of cholesteric light scattering \cite{deGennes}).  The tilt mode is visualized in Fig. 8c; the polarization mode is not visualized.

A coupling between tilt and polarization (even if weak -- i.e., small $m_{23}$) has an important physical significance.  If an electric field is applied in the $y$-direction, it induces a polarization $\delta p_y$.  Because of the coupling, it must also induce a tilt $t_x$.  Hence, the $\mathrm{N_{TB}}$ phase has an electroclinic effect, analogous to a chiral smectic-A phase.  The sign of the electroclinic effect depends on the sign of $m_{23}$, which is controlled by the sign of the helicity $q_0$.  For that reason, domains of right- and left-handed helicity must have opposite electroclinic effects.  In earlier work, a weak electroclinic effect was observed experimentally and modeled by a different theoretical method \cite{CMeyer_PRL}.  Here, we see that it is a consequence of the coarse-grained free energy.

(4, 5) The coarse-grained director tilt $t_y$ and polarization $\delta p_x$ form another pair of nonhydrodynamic normal modes, which is degenerate with the previous pair.

(6) The polarization component $\delta p_z$ is itself a nonhydrodynamic normal mode.  Its relaxation rate is
\begin{equation}
\Gamma^{p^\prime} = m_{66}/\gamma_{p^\prime},
\label{Gammapprimeintwistbend}
\end{equation}
where $\gamma_{p^\prime}$ is the viscosity of this mode.

If the wavevector $\mathbf{q}$ is small but nonzero, the five nonhydrodynamic modes are only slightly changed.  To model their relaxation rates, we can still use Eqs. (\ref{GammatGammapintwistbend}) and (\ref{Gammapprimeintwistbend}) derived above.  However, the hydrodynamic mode is more significantly changed.  We can consider the cases of $\mathbf{q}$ parallel and perpendicular to the $z$-direction separately:

For $\mathbf{q}$ in the $z$-direction, the hydrodynamic mode still involves the phase $\phi$ by itself, not coupled with any other coarse-grained degrees of freedom.  This mode is visualized in Fig.~8d.  It is a $z$-dependent rotation of the heliconical director field $\hat{\mathbf{n}}(\mathbf{r})$, which does not change the coarse-grained director $\hat{\mathbf{t}}$.  Equivalently, this mode can be regarded as a $z$-dependent displacement $u=-\phi/q_0$ of the pseudo-layers, leading to alternating compression and dilation of the pseudo-layer structure.  In the limit of long wavelength (small $\mathbf{q}$), the free energy cost of this fluctuation is $\frac{1}{2}B_\text{eff}q_z^2 |u_\mathbf{q}|^2$, where
\begin{equation}
B_\text{eff} = \left(K_2 \sin^4 \beta + K_3 \sin^2 \beta \cos^2 \beta + \kappa  p_0^2 \right) q_0^2 .
\end{equation}
Hence, the relaxation rate is $\Gamma^u (q_z) = \frac{1}{2}\gamma_u^{-1} B_\text{eff}q_z^2$, where $\gamma_u$ is the relevant viscosity.

For $\mathbf{q}$ in the $x$-direction, the hydrodynamic normal mode is a linear combination of $\phi$, $t_x$, and $\delta p_y$, as visualized in Fig.~8e.  This mode is an $x$-dependent rotation of the $\hat{\mathbf{n}}(\mathbf{r})$, or equivalently an $x$-dependent displacement of the pseudo-layers, leading to curvature of the pseudo-layer structure.  This displacement is accompanied by a tilt of the coarse-grained director in the $x$-direction, so that the local $\hat{\mathbf{t}}$ remains normal to the local pseudo-layers.  If $\mathbf{q}$ is in any other direction in the $(x,y)$ plane, the same description applies with the corresponding rotation.  The free energy cost of this fluctuation is
$\frac{1}{2}K_\text{eff}q_\perp^4 |u_\mathbf{q}|^2$, where
\begin{equation}
K_\text{eff} = K_1 \cos^2 \beta + \frac{1}{2} K_3 \sin^2 \beta + \frac{1}{2} \kappa p_0^2,
\end{equation}
to lowest order in small $\beta$.  Hence, the relaxation rate is $\Gamma^u (q_\perp) = \frac{1}{2}\gamma_u^{-1} K_\text{eff}q_\perp^4$.

In both cases, the effective elasticity of the $\mathrm{N_{TB}}$ phase is equivalent to a smectic-A phase, with $B_\text{eff}$ and $K_\text{eff}$ playing the roles of the elastic moduli for compression and bending of the smectic layers, respectively.  In that way, our coarse-graining of the $\mathrm{N_{TB}}$ phase is analogous to earlier work on coarse-graining of the cholesteric phase, which also has effective smectic elasticity \cite{Lubensky}.

\section{Discussion}
We can now compare the calculated normal modes with the light scattering experiment.

\subsection{Nematic phase}
The fluctuating part of the dielectric tensor can be expressed in terms of the normal modes using the relation
\begin{equation}
\epsilon_{ij}(\mathbf{r}) = \Delta \epsilon^n n_i n_j + \Delta \epsilon_{sat}^p p_i p_j
\label{dielectric}
\end{equation}
where $(i,j) = (x,y,z)$, $\Delta \epsilon^n$ is the dielectric anisotropy associated with the orientational ordering of $\hat{\mathbf{n}}$, and $\Delta \epsilon_{sat}^p$ is the saturated value of the dielectric anisotropy associated with the $\mathbf{p}$ ordering.

In geometry G1, with $q_z=0$, the fluctuations in $\hat{\mathbf{n}}$ and $\mathbf{p}$ decouple. The former yield the usual pair of hydrodynamic director modes ($n_1$, $n_2$), while the latter produce a doubly degenerate nonhydrodynamic mode associated with $p_x$, $p_y$, plus an independent nonhydrodynamic mode associated with $p_z$. Assuming large coefficient $\eta$ in Eq.~(1), we can neglect $p_z$. Since the incident polarization in geometry G1 is along $\hat{\mathbf{z}}$, the relevant elements of $\epsilon_{ij}$ for depolarized scattering are $\epsilon_{xz}$ and $\epsilon_{yz}$. Assuming negligible $p_z$, these elements are dominated by the director modes, and specifically in our experiment for large $\theta_s$, by the splay fluctuations in the normal mode $n_1$. Therefore, in agreement with our experimental results for geometry G1, the model with large $\eta$ predicts that the DLS correlation function is described by a single exponential decay (with relaxation rate $\Gamma_1^n$), and that the contribution from nonhydrodynamic polarization fluctuations is not observable.

The situation is different in geometry G2, where $\mathbf{q} = q_x \hat{\mathbf{x}} + q_z \hat{\mathbf{z}}$ for scattering in the $x$-$z$ plane. (The choice of $x$-$z$ or $y$-$z$ is arbitrary.) In depolarized DLS, with the incident light polarized along $\hat{\mathbf{y}}$, we probe fluctuations $\epsilon_{zy}$ and $\epsilon_{xy}$. From the former, we expect and observe the $n_2$ (twist-bend) hydrodynamic mode. The latter ($\epsilon_{xy}$) couples to nonhydrodynamic polarization fluctuations transverse to the nematic ordering axis, which contribute maximally to the DLS signal in the ``dark director" limit of G2, where $\epsilon_{zy} \rightarrow 0$ and $\epsilon_{xy}$ dominates. 

Since $\epsilon_{xy}$ is quadratic in $\mathbf{p}$ fluctuations (Eq.~(\ref{dielectric})), DLS probes the higher-order time correlation function $C(p_x,p_y) = \langle p_x^\ast (0) p_y^\ast (0) p_x (\tau) p_y (\tau) \rangle$ ($\tau$ = delay time). Based on the normal mode structure of the free energy for $q_z \neq 0$ and assuming the fluctuations are Gaussian random variables with zero mean, $C(p_x,p_y)$ can be reduced to $\langle p_x^\ast(0) p_x(\tau) \rangle \langle p_y^\ast(0) p_y(\tau) \rangle$. The normal modes are linear combinations of $n_1, p_x$ and of $n_2, p_y$, yielding for $K_1 \approx K_2$ a pair of nearly degenerate hydrodynamic modes and a pair of nearly degenerate nonhydrodynamic modes. 

In the limit that the energy associated with $\mathbf{p}$ fluctuations is much greater than that of the $\mathbf{p}-\hat{\mathbf{n}}$ coupling, and that the latter is much greater than the elastic energy of $\hat{\mathbf{n}}$ fluctuations, the correlation function is a double exponential decay, as observed in our experiment \cite{Rosenblatt_PRE}, with the faster decay characterized by relaxation rate $\Gamma_2^p \sim$~(constant in $q$) for the $\mathbf{p}$ fluctuations, and the slower characterized by a rate $\Gamma_2^n \sim q^2$ representing a mixture of director modes. This mixture could explain the broadening of the slower decay indicated by our data analysis. 

Outside of the ``dark director" geometry, the twist-bend director scattering from $\epsilon_{zy}$, which is linear in $n_2$, prevails, and the fast decay makes only a weak contribution to the DLS correlation function -- again in agreement with the experiment. The relaxation rate of the director mode (Fig.~6) decreases as $T \rightarrow T_{TB}$ from above, but only by a factor $\sim 1.6$. This modest decrease remains consistent with the expected softening of bend fluctuations, $K_3^\text{eff} \rightarrow 0$ as $T \rightarrow T_{TB}$ [see Eqs.~(11) and (7)], since Eq.~(12) indicates that the relaxation rate $\Gamma_2^n \simeq K_2 q_\perp^2 / \gamma_n$ for the condition $q_\perp^2 \gg q_z^2$ in geometry G2. Thus, in the scattering geometry used, $\Gamma_2^n$ is not very sensitive to the temperature dependence of $K_3^\text{eff}$.

\subsection{Twist-bend phase}
The spectrum of modes is related to fluctuations of the dielectric tensor through a modified version of Eq.~(\ref{dielectric}), where $\Delta \epsilon^n n_i n_j$ is replaced by $\Delta \epsilon^t t_i t_j$.  The hydrodynamic mode is the extension into the $\mathrm{N_{TB}}$ phase of the splay-bend director mode $n_1$, which is observed in the nematic phase in geometry G1.  The nonhydrodynamic tilt mode is the extension of the hydrodynamic twist-bend director mode $n_2$, which is observed in the nematic phase in geometry G2. It acquires a large energy gap when the heliconical structure forms, analogous to the gap in $n_2$ that develops in a smectic-A phase due to the large energy cost of tilting the director away from the layer normal.  The coarse-grained model thus accounts for both the slow hydrodynamic mode (data labeled (a) in the bottom panel of Fig.~4) and the slower of the pair of nonhydrodynamic modes (data labeled (b) in the bottom panel of Fig.~4), which are observed in experimental geometries G1 and G2, respectively.

The faster nonhydrodynamic mode in the $\mathrm{N_{TB}}$ phase is detected in the ``dark director" limit of geometry G2 (see correlation data labeled (c) in the bottom panel of Fig.~4).  As in the nematic case, it can be associated with fluctuations of the polarization $(\delta p_x, \delta p_y)$.  Because the polarization fluctuations are only observed for a scattering geometry where the $\hat{\mathbf{t}}$ fluctuations are ``dark," the coupling between tilt and polarization fluctuations must be weak.

The coarse-grained theory predicts additional terms in the expression for the energy gap of these fluctuations in the $\mathrm{N_{TB}}$ phase compared with the nematic phase [see Eq.~(\ref{GammatGammapintwistbend}b) compared with (\ref{Gammapinnematic})]. These terms imply an increase in the relaxation rate $\Gamma_p$ of the polarization mode at $T_{TB}$, which is consistent with the experimentally observed behavior (Fig.~6). According to the model, this increase in $\Gamma_p$ signals a transition to a heliconical structure with $\beta\neq 0$ and $p_0 \neq 0$.

The slow relaxation process that mixes with the fast polarization fluctuations in the correlation function is also explained by the theory: When $q_z$ and $q_\perp$ are non-zero, as is generally the case in the G2 geometry, $\delta p_x$ and $\delta p_y$ mix with the slow hydrodynamic variable $\phi$ and with $\hat{\mathbf{t}}$, and thus the correlation function contains a slow component corresponding to undulation of the pseudo-layers and splay of $\hat{\mathbf{t}}$. 

The final nonhydrodynamic mode predicted by the theory, related to $\delta p_z$, has an even higher relaxation rate, which is not detected in our experiment.  This high relaxation rate implies a relatively large value for the coefficient $\eta$ in the free energy of Eq.~(\ref{fntb}).

To fit the experimental data for relaxation rates as functions of temperature, we combine Eq.~(\ref{Gammapinnematic}) in the nematic phase and Eqs.~(\ref{matrixelements}--\ref{GammatGammapintwistbend}) in the $\mathrm{N_{TB}}$ phase.  For the equilibrium cone angle $\beta$ and pitch wavenumber $q_0$, we use the leading terms in Eqs.~(\ref{q0min}--\ref{betamin}) near the second-order transition, which give $\sin^2 \beta \approx p_0 (\kappa/K_2)^{1/2}$ and $q_0 \approx (\Lambda/K_3) p_0^{1/2} (K_2/\kappa)^{1/4}$.  For the equilibrium polarization $p_0$, we use the approximation of Eq.~(\ref{p0scaling2}), derived with the assumption of small polarization elasticity $\kappa$.  The predicted relaxation rates then become
\begin{subequations}
\label{Gammatocompare}
\begin{align}
& \Gamma^p (T>T_{TB})= \frac{\mu_0}{\gamma_p}\left[\frac{\Lambda^2}{K_3 \mu_0}+ (T-T_{TB})\right], \\
& \Gamma^t (T<T_{TB}) = \frac{\Lambda^2 (K_1 + K_2)\mu_0 (T_{TB}-T)}{2 \gamma_t K_3^2 \nu}, \\
& \Gamma^p (T<T_{TB}) = \frac{\mu_0}{\gamma_p}\Biggl[\frac{\Lambda^2}{K_3 \mu_0}+\eta\sqrt{\frac{\kappa(T_{TB}-T)}{K_2 \mu_0 \nu}}\nonumber\\
& \quad\quad\quad\quad\quad\quad\quad\quad\quad +(T_{TB}-T)\Biggr].
\end{align}
\end{subequations}
We can compare Eqs.~(\ref{Gammatocompare}) directly with the data in Fig.~6.  In this comparison, we assume that the orientational viscosities $\gamma_t$ and $\gamma_p$ do not vary strongly with temperature.

First, fitting Eq.~(\ref{Gammatocompare}a) to the data for $\Gamma^p$ in the nematic phase, we find $\mu_0/\gamma_p = 3600$~s$^{-1}$~K$^{-1}$ and $\Lambda^2/(K_3 \mu_0) = 30$~K. The fit is shown as a solid line in Fig.~6 (bottom panel, $T>T_{TB}$).

Second, the data for $\Gamma^t$ in the $\mathrm{N_{TB}}$ phase are consistent with the linear dependence in Eq.~(\ref{Gammatocompare}b). This consistency confirms that the experiment is in the regime where $p_0$ follows the the approximation of Eq.~(\ref{p0scaling2}) rather than Eq.~(\ref{p0scaling1}). The experimental slope corresponds to the combination of parameters $\Lambda^2 (K_1 + K_2)\mu_0/(2 \gamma_t K_3^2 \nu)=84000$~s$^{-1}$~K$^{-1}$. This fit is shown as a solid line in Fig.~6 (top panel, $T<T_{TB}$).

Third, the data for $\Gamma^p$ in the $\mathrm{N_{TB}}$ phase can be fit to the expression in Eq.~(\ref{Gammatocompare}c), as shown by the solid line in Fig.~6 (bottom panel, $T<T_{TB}$). In this fit, we use the parameters $\mu_0/\gamma_p = 3600$~s$^{-1}$~K$^{-1}$and $\Lambda^2/(K_3 \mu_0) = 30$~K obtained from the analysis of $\Gamma^p$ in the nematic phase. The fit yields $\eta\kappa^{1/2}(K_2 \mu_0 \nu)^{-1/2}= 1200$~K$^{1/2}$.

We now combine the last fit result with two estimates.  From the argument after Eq.~(\ref{betamin}), we have $(\kappa/K_2)^{1/2}\simeq 0.03$.  Furthermore, if we use Eq.~(\ref{p0scaling2}) and take $p_0 \simeq 0.1$ at $T_{TB}-T = 1$~K, we find $(\mu_0/\nu)^{1/2} \simeq 0.1$~K$^{-1/2}$.  Together with the fit result, these estimates give $\eta/\mu_0 \simeq 4.0 \times 10^5$~K. This large value indicates that the relaxation rate $\Gamma^{p^\prime}$ of longitudinal polarization fluctuations in Eqs.~(\ref{Gammapprimeinnematic}) and (\ref{Gammapprimeintwistbend}) is much larger than $\Gamma^p$, and hence explains why those fluctuations are not observed in our experiment.

We may also verify two conditions on which our analysis is predicated: (1) that $T_{TB}-T > \Delta T_x = 9\Lambda^4 \kappa K_2/(4 K_3^4 \mu_0 \nu)$ (see Eq.~(10) and accompanying discussion above) over the temperature range of our data in the $\mathrm{N_{TB}}$ phase, meaning Eq.~(9) applies, and therefore Eq.~(29b) is valid; and (2) that $m_{22} m_{33} \gg m_{23}^2$, which validates the decoupling approximation for the polarization and tilt modes, and hence the use of Eqs.~(29a) and (29b) for their relaxation rates. 

First, using the numerical results above, we find that $\Delta T_x \simeq 0.02$~K, and thus confirm that our $\mathrm{N_{TB}}$ data are strictly in the regime $T_{TB}-T \gg \Delta T_x$ where Eq.~(9) applies. Next, from Eqs.~(4) and (5) for small $\sqrt{\kappa/K_2}$, we have $\sin^2 \beta \simeq p_0 \sqrt{\kappa/K_2}$ and $q_0 \simeq \sqrt{p_0} (\Lambda/K_3) (K_2/\kappa)^{1/4}$. Eqs.~(23) then imply $m_{22} \simeq (\Lambda p_0/K_3)^2 (K_1+K_2)/2$, $m_{33} > \eta p_0 \sqrt{\kappa/K_2}$, and $m_{23} = - (\Lambda^2/2 K_3)(\kappa/K_2)^{1/4} p_0^{3/2}$. Thus, the condition $m_{22} m_{33} \gg m_{23}^2$ may be written as $\eta/\mu_0 \gg (\Lambda^2/(2 K_3 \mu_0))(K_3/(K_1+K_2))$. Using the parameters from the fits given above, this reduces to $4 \times 10^5$~K$\gg 15$~K$(K_3/(K_1+K_2))$, which is clearly valid. 


Finally, consider the data for the inverse scattering intensity $I_2^{-1}$ in Fig.~7.  These data were recorded in geometry G2 for $\theta_i = 15^\circ$, $\theta_s = 40^\circ$, where the scattering is dominated by optic axis fluctuations (i.e., $\hat{\mathbf{n}}$ or $\hat{\mathbf{t}}$).  In each phase, $I_2^{-1}$ is proportional to the free energy density of these fluctuations.  On this basis, we can make two useful comparisons between experiment and theory: 

(1) Since $I_2^{-1} \propto \gamma_t \Gamma^t$ in the $\mathrm{N_{TB}}$ phase, and since $\Gamma^t$ is essentially linear in $T_{TB}-T$ (Fig.~6), we expect and observe the same for $I_2^{-1}$ (Fig.~7).

(2) In the nematic phase, the free energy density of director fluctuations is given by $\frac{1}{2} K_2 q_\perp^2$ (from the Frank free energy with the experimental condition $K_2 q_\perp^2 \gg K_3 q_z^2$ appropriate for geometry G2).  In the $\mathrm{N_{TB}}$ phase, the free energy density of coarse-grained director fluctuations is given by $\frac{1}{2} (K_1 + K_2)q_0^2 \sin^2 \beta$ [from Eq.~(\ref{GammatGammapintwistbend}a) for $\gamma_t \Gamma^t$ combined with the result $p_0 \approx (K_3/\Lambda) q_0 \sin \beta$ near the transition].  Hence, the ratio of scattering intensities in the two phases should be
\begin{equation}
\frac{I_2(T>T_{TB})}{I_2(T<T_{TB})}\approx\frac{(K_1 + K_2) q_0^2 \sin^2 \beta}{2 K_2 q_\perp^2}\approx\frac{q_0^2 \sin^2 \beta}{q_\perp^2} \nonumber.
\end{equation}
From Ref.~\cite{Borshch_Nature}, using relative values of the optical birefringence at $T = T_{TB}$ and $T-T_{TB} = -5^\circ$C, we estimate $\beta = 7.5^\circ$.  From the same reference, FFTEM textures show that the pitch is $2\pi/q_0 = 9.3$~nm.  In our experimental geometry, $q_\perp = 2 \pi (\sin \theta_i + \sin \theta_s)/\lambda = 0.011$~nm$^{-1}$.  Combining these numbers gives
\begin{equation}
\frac{I_2(T>T_{TB})}{I_2(T<T_{TB})}\approx 70 \nonumber.
\end{equation}
By comparison, the experimental intensity ratio in Fig.~7 (between the nematic phase just above the transition and the $\mathrm{N_{TB}}$ in the middle of its range, 5$^\circ$C below the transition) is approximately 60.  This quantitative similarity gives additional support to the theory.

\section{Conclusion}
Our DLS study of a twist-bend nematic liquid crystal demonstrates the presence of a pair of temperature-dependent, nonhydrodynamic fluctuation modes connected to the $\mathrm{N_{TB}}$ structure. One of these modes is associated with twist-bend director fluctuations in the presence of a short-pitch heliconical modulation of $\hat{\mathbf{n}}$, while the other is accounted for by fluctuations in a vector order parameter that corresponds to a helical polarization field coupled to the director modulation. The behavior of both modes, as well as the presence of a single hydrodynamic mode in the $\mathrm{N_{TB}}$ phase (associated with splay fluctuations of the helical pitch axis), are quantitatively explained by a theoretical model based on two components: (1) a Landau-de Gennes free energy density, which is expanded in the director and polarization fields; and (2) A ``coarse-graining" of this free energy that maps the heliconical structure onto a smectic-like system characterized by a ``pseudo-layer" displacement field and an effective director normal to the layers. This model predicts one hydrodynamic and one non-hydrodynamic ``layer"-director mode, and also reveals how the distortions of the pseudo-layers couple to fluctuations in the polarization field.

It will be interesting to test this mapping further -- for example, by designing experiments to determine the magnitude of the effective elastic constant for layer compression as a function of heliconical pitch \cite{Bconstant}. It could also be illuminating to probe the response of the polarization mode to an applied electric field. Finally, extending the Landau-deGennes theory to include a first-order N-$\mathrm{N_{TB}}$ transition may prove useful for understanding experimental results on a wider range of dimers or monomer/dimer mixtures.  

\begin{acknowledgments}
We are grateful to S. Pardaev for assistance in collecting the light scattering data, and to A. R. Baldwin for his technical expertise in developing the correlation hardware and software used in making the measurements. We thank the agencies for support: the NSF under grants DMR-1307674 (ZP, JG, AJ, SS), DMR-1409658 (SMS, JVS), and DMR-1410378 (VB, ODL); the EPSRC under grant EP/J004480/1 (GM and CW); and the EU under project 216025 (MGT).
\end{acknowledgments}


\begin{references}
\bibitem{Chen_PNAS} D. Chen, J. H. Porada, J. B. Hooper, A. Klittnick, Y. Shen, M. R. Tuchband, E. Korblova, D. Bedrov, D. M. Walba, M. A. Glaser, J. E. Maclennan, and N. A. Clark, Proc. Natl. Acad. Sci. USA {\bf 110}, 15931 (2013).
\bibitem{Meyer} R. B. Meyer in {\it Molecular Fluids. Les Houches Lectures, 1973}, R. Balian and G. Weill, eds (Gordon and Breach, 1976), pp. 271-343.
\bibitem{Dozov} I. Dozov, Europhys. Lett. {\bf 56}, 247 (2001).
\bibitem{Cestari_PRE} M. Cestari, S. Diez-Berart, D. A. Dunmur, A. Ferrarini, M. R. de la Fuente, D. J. B. Jackson, D. O. Lopez, G. R. Luckhurst, M. A. Perez-Jubindo, R. M. Richardson, J. Salud, B. A. Timimi, and H. Zimmermann, Phys. Rev. E {\bf 84}, 031704 (2011).
\bibitem{Borshch_Nature} V. Borshch, Y.-K. Kim, J. Xiang, M. Gao, A. Jakli, V. P. Panov, J. K. Vij, C. T. Imrie, M. G. Tamba, G. H. Mehl, and O. D. Lavrentovich, Nat. Commun. {\bf 4}, 2635 (2013).
\bibitem{Luckhurst} A. Ferrarini, G. R. Luckhurst, P. L. Nordio, and S. J. Roskilly, Chem. Phys. Lett. {\bf 214}, 409 (1993); J. Chem. Phys. {\bf 100}, 1460 (1994).
\bibitem{Kamien} A. B. Harris, R. D. Kamien, and T. C. Lubensky, Rev. Mod. Phys. {\bf 71}, 1745 (1999).
\bibitem{BMeyer} R. B. Meyer, Phys. Rev. Lett. {\bf 22}, 918 (1969).
\bibitem{CMeyer_PRL} C. Meyer, G. R. Luckhurst, and I. Dozov, Phys. Rev. Lett. {\bf 111}, 067801 (2013).
\bibitem{Shamid_PRE} S. M. Shamid, S. Dhakal, and J. V. Selinger, Phys. Rev. E {\bf 87}, 052503 (2013).
\bibitem{Adlem_PRE}K. Adlem, M. Copic, G. R. Luckhurst, A. Mertelj, O. Parri, R. M. Richardson, B. D. Snow, B. A. Timimi, R. P. Tuffin, and D. Wilkes, Phys. Rev. E {\bf 88}, 022503 (2013).
\bibitem{Lubensky} T. C. Lubensky, Phys. Rev. A {\bf 6}, 452 (1972); L. Radzihovsky and T. C. Lubensky, Phys. Rev. E {\bf 83}, 051701 (2011).
\bibitem{Challa_PRE} P. K. Challa, O. Parri, C. T. Imrie, S. Sprunt, J. T. Gleeson, O. Lavrentovich, and A. Jakli, Phys. Rev. E {\bf 89}, 060501(R) (2014).
\bibitem{Salili_RSC} S. M. Salili, C. Kim, S. Sprunt, J. T. Gleeson, O. Parri, and A. Jakli, RSC Adv {\bf 4}, 57419 (2014).
\bibitem{CMeyer_SoftMatter}  C. Meyer and I. Dozov, Soft Matter, 2016, DOI: 10.1039/C5SM02018B.
\bibitem{Mehl} V. P. Panov, M. Nagaraj, J. K. Vij, Y. P. Panarin, A. Kohlmeier, M. G. Tamba, R. A. Lewis, and G. H. Mehl, Phys. Rev. Lett. {\bf 105}, 167801 (2010).
\bibitem{Lavrentovich_PRL} Z. Li and O.D. Lavrentovich, Phys. Rev. Lett. {\bf 73}, 280 (1994).
\bibitem{Jakli_APL} A. Jakli and A. Saupe, Appl. Phys. Lett. {\bf 65}, 2777 (1994).
\bibitem{hyd_mode} The temperature dependence of the hydrodynamic director modes on the nematic side has previously been explored on a different dimer material -- see Ref.~\cite{Adlem_PRE}. Results for this dependence on the mixture studied in the present work are similar to those presented in this reference.
\bibitem{Rosenblatt_PRE} Interestingly, a double exponential decay of the DLS correlation function, with a nonhydrodynamic fast component, was observed many years ago in an ordinary rodlike nematic (8CB) [see S. Tripathi, H. Zhong, R. G. Petschek, and C. Rosenblatt, Phys. Rev. E {\bf 52},5004 (1995)]. These authors ultimately attributed the fast process (which had a temperature-independent relaxation rate comparable to the rate we observe) to biaxial correlations among molecules and/or to molecular conformational fluctuations with a large biaxial character. In the case of the bent-shaped LC dimers, one can imagine a natural connection between conformational motions and a polar (vector) order parameter that drives the N to $\mathrm{N_{TB}}$ transition. Polar correlations in the plane perpendicular to the main director possess a biaxial charater, but are characterized by a vector rather than second rank tensor order parameter.
\bibitem{deGennes} P.-G. de Gennes and J. Prost, {\it The Physics of Liquid Crystals}, second edition (Oxford, 1993), section 6.3.1.3.
\bibitem{Bconstant} In fact, using the values of cone angle $\beta$ and pitch wavenumber $q_0$ from the Discussion section, assuming $K_2$ and the ``bare" $K_3$ (maximum value in the nematic phase) are in the pN range (see ref.~\cite{Borshch_Nature}), and neglecting the contribution of $\kappa$ (polarization elastic constant), one estimates from Eq.~(26) that $B_\text{eff}$ should be $\sim 100$ times lower in the $\mathrm{N_{TB}}$ phase than in the smectic-A phase of a typical thermotropic liquid crystal. This estimate agrees with recent experimental results in ref.~\cite{Salili_RSC}.
\end{references}
\end{document}